\documentclass[11pt]{scrartcl} 
\usepackage[letterpaper, margin=0.8in,bottom=1in]{geometry}
\usepackage[utf8]{inputenc}
\usepackage{amsmath}
\usepackage{mathrsfs}
\usepackage{mathtools}
\usepackage{bigints}
\usepackage{textgreek}
\usepackage{graphicx} 
\graphicspath{{figures/}} 
\usepackage[font=small,labelfont=bf]{caption}
\setcapindent{0em}
\usepackage{authblk}
\usepackage{tabularx}
\usepackage{booktabs} 
\usepackage{makecell} 
\usepackage{multirow} 
\usepackage{xcolor}
\usepackage[super,numbers,sort,compress]{natbib}

\usepackage[colorlinks,allcolors=cyan!70!black]{hyperref} 
\usepackage{scalerel} 
\usepackage{lineno}

\newcounter{suppnote}

\newcommand{\supplementarynote}[2]{%
  \clearpage
  \refstepcounter{suppnote}%
  \section*{Supplementary Note \thesuppnote: #1}%
  \addcontentsline{toc}{section}{Supplementary Note \thesuppnote: #1}%
  \label{#2}%
}

\newcommand{\suppnoteref}[1]{%
  \hyperref[#1]{Supplementary Note \ref*{#1}}%
}

\title{ \textbf{Dealloying by peritectic melting}  }

\author[]{Mingwang Zhong}
\author[1,*]{Alain Karma}
\affil[]{\small Physics Department and Center for Interdisciplinary Research in Complex Systems\\

Northeastern University, Boston, Massachusetts, 02115, USA }
\affil[*]{email: a.karma@northeastern.edu}

\date{}

\begin{document}

\maketitle

\section*{Abstract}

Peritectic melting of Ti--Ag has been shown experimentally to form bicontinuous structures, but the mechanism remains unclear. Here we use phase-field simulations to show that these structures arise from a morphological instability of liquid film migration in three dimensions: a Ti-rich solid growing through an Ag-rich liquid film develops a branched seaweed or dendritic structure whose side branches coalesce to form handles, generating a high-genus bicontinuous topology. A sharp-interface theory predicts a solidification-front velocity and an initial ligament width that are constant in time, in contrast to liquid metal dealloying; subsequent $t^{1/3}$ coarsening reproduces the experimentally observed final ligament width.

\section*{Impact statement}
Computational modeling reveals the interfacial pattern formation mechanism that transforms a homogeneous peritectic alloy into a two-phase bicontinuous structure by unstable liquid film migration and interface coalescence during the melting process.

\section*{Keywords}
peritectic melting; dealloying; liquid film migration; bicontinuous structure; phase-field simulation

\section*{Introduction}

Dealloying is a powerful route for fabricating bicontinuous metals with large interfacial area and useful functional and mechanical properties. These structures enable applications in energy storage \cite{guo2008nanostructured, shao2020nanoporous}, catalysis \cite{zielasek2006gold, wittstock2010nanoporous}, sensing \cite{hu2008electrochemical, zhang2012nanoporous}, and nanomechanics \cite{hodge2007scaling, jin2009deforming}. Conventional electrochemical dealloying (ECD) selectively dissolves less noble elements from a precursor alloy into an electrolyte \cite{erlebacher2001evolution, hayes2006monolithic, snyder2010oxygen,mccue2018pattern}, while liquid metal dealloying (LMD) uses a molten metal to dissolve the miscible elements and extends bicontinuous structure formation to systems such as Ta--Ti, Fe--Cr--Ni, and Si--Mg \cite{harrison1959attack, wada2011dealloying, geslin2015topology, wada2014bulk, kim2015optimizing, mccue2016size, greenidge2020porous, lai2022topological, lai2022microstructural, song2022ultrafine}. Vapor phase dealloying (VPD) further exploits vapor pressure differences to selectively remove elements into a vacuum \cite{lu2018three, han2019vapor, lu2021vapor}. In all these cases, selective removal or dissolution into an external medium drives the formation of an interconnected solid/liquid or solid/vapor morphology.

Recent experiments have shown that isothermal melting of a peritectic $\text{Ti}_{50}\text{Ag}_{50}$ parent $\beta$ alloy can also generate bicontinuous microstructures \cite{hu2019evolution, li2025peritectic}. This process differs fundamentally from ECD, LMD, and VPD because it does not require an external dealloying medium. Upon heating above the peritectic temperature $T_p$, the metastable $\beta$ phase decomposes into a Ti-rich $\alpha$ solid and an Ag-rich liquid. Redistribution of Ti and Ag is therefore supported internally by diffusion within the newly formed liquid phase. Among conventional dealloying routes, LMD provides the closest comparison to peritectic melting (PM) because both processes involve two solid phases and a liquid metal. However, in LMD the liquid is an external reservoir, and solute must diffuse through the growing dealloyed layer to the surrounding melt. This produces time-dependent dealloying kinetics, with the dealloying front velocity $v\propto 1/\sqrt{t}$ and the ligament width at the front $\lambda_0\sim 1/\sqrt{v}$ decreasing and increasing in time, respectively \cite{geslin2015topology,lai2022topological}. In PM, by contrast, the liquid forms locally at the melting front, suggesting that the dealloying kinetics may differ from LMD. However, the mechanism by which PM produces a bicontinuous structure and the scaling laws governing $v$ and $\lambda_0$ as a function of time and driving force (superheating $\Delta T=T-T_p$) remain unknown. 

PM has traditionally been associated with liquid film migration (LFM), in which a thin Ag-rich liquid film advances into the parent $\beta$ alloy while the $\alpha$--liquid and $\beta$--liquid interfaces migrate in concert \cite{muschik1989melting,boussinot2010kinetics}. Viewed as a smooth, uniformly advancing front, LFM is topology-preserving and should not by itself generate a high-genus bicontinuous structure.
However, recent electron microscopy observations of Ti–Ag peritectic melting suggest that the bicontinuous α/L network develops through LFM \cite{li2025peritectic}: an $\alpha$ nucleus formed at a premelted $\beta$--$\beta$ grain boundary remains separated from the neighboring $\beta$ grain by an Ag-rich liquid film and propagates with the receding $\beta$--liquid front. This raises the central question: how can LFM, apparently a planar film-migration process, generate high-genus topology?

Here we use phase-field (PF) simulations of PM to answer this question. 
To isolate the LFM mechanism from the $\alpha$--$\beta$--liquid triple-junction geometry, we initiate growth from a Ti-rich $\alpha$ nucleus fully immersed in the Ag-rich liquid film between neighboring $\beta$
grains. This avoids imposing uncertain contact angles determined by the interfacial free energies $\gamma_{l\alpha}$,
$\gamma_{l\beta}$, and $\gamma_{\alpha\beta}$, which are not accurately known for Ti–Ag, while preserving the essential feature of LFM: two solid–liquid interfaces migrating together across a thin liquid layer.
Seeding growth heterogeneously from an $\alpha$ nucleus on a $\beta$
grain, as observed experimentally \cite{hu2019evolution, li2025peritectic}, instead of homogeneously inside the inter-grain liquid film yields the same large-scale morphological development of the $\alpha$--liquid interface. We therefore focus on the homogeneous case, which isolates the mechanism of bicontinuous structure formation from the details of initial nucleus spreading that depend on the triple-junction geometry.

Since crystalline anisotropy is well-known to influence solidification morphologies \cite{barbieri1989predictions,haxhimali2006orientation}, 
we consider both cases where $\gamma_{l\alpha}$ is isotropic and weakly anisotropic. The simulations reveal that, in both two and three dimensions (2D and 3D), the $\alpha$--liquid interface becomes morphologically unstable forming branched seaweed or dendritic structures with the topology evolving in 3D by coalescence of neighboring branches that form handles and produce a high-genus $\alpha/L$ network. They further show that, in contrast to LMD, $v$ and $\lambda_0$ are constant in time, and the structure coarsens behind the front with $\lambda \sim t^{1/3}$ as in Ostwald ripening \cite{lifshitz1961kinetics,wagner1961theory} instead of a smaller coarsening exponent varying between 1/4 and 1/3 due to a mix of surface and bulk diffusion in LMD \cite{mccue2018pattern,lai2022topological}.
We extend existing sharp-interface theories of dendritic solidification \cite{ivantsov1947temperature,barbieri1989predictions} 
and LFM \cite{brener2005velocity,brener2007melting,boussinot2010kinetics} to derive the scaling laws $v\sim \Delta T^2$ and $\lambda_0\sim 1/\Delta T$ that are validated by simulations.
These results identify branching and coalescence within morphologically unstable LFM as the mechanism of bicontinuous structure formation, establishing PM as a dealloying pathway distinct from LMD.

\section*{Methods}

We simulated peritectic melting of Ti--Ag using a multiPF model adapted from a quantitative formulation for eutectic solidification~\cite{folch2005quantitative}. The Ag-rich liquid, Ti-rich $\alpha$ solid, and parent $\beta$ alloy are represented by phase fields $\phi_l$, $\phi_\alpha$, and $\phi_\beta$, constrained by $\phi_l+\phi_\alpha+\phi_\beta=1$, together with a conserved Ag concentration field $c$. The free-energy functional contains gradient, triple-well, and concentration-coupling terms calibrated to the linearized Ti--Ag phase diagram and to the interfacial free energies $\gamma_{l\alpha}$, $\gamma_{l\beta}$, and $\gamma_{\alpha\beta}$. The phase fields evolve by dissipative relaxation subject to the PF constraint, while $c$ evolves by diffusion with mobility restricted to the liquid phase, so that solute transport vanishes in the solids and reduces to Fickian diffusion in the bulk liquid.

To examine the effect of crystalline anisotropy, the $\alpha$--liquid interfacial free energy was taken to be either isotropic or weakly anisotropic with cubic symmetry \cite{barbieri1989predictions,haxhimali2006orientation}. The equations were solved on a uniform Cartesian grid using second-order central differences and explicit time integration in a CUDA C/C++ implementation. Topological analysis of the dealloyed region was performed by reconstructing the $\alpha$--liquid interface using marching cubes~\cite{lorensen1998marching}; the genus $g$ and ligament size $\lambda$ were then obtained from the triangulated surface and structure-factor analysis. Full model equations, numerical parameters, calibration details, anisotropy implementation, and topology-analysis procedures are provided in \textcolor{cyan!70!black}{Supplementary Notes 1--3}, with parameters listed in Supplementary Table~\ref{tabs:constant} \cite{dinsdale1991sgte, li2005experimental, zhong2025quantification, tourret2015growth}.

\begin{figure*}[!hb]
\centering
\includegraphics[width=\textwidth]{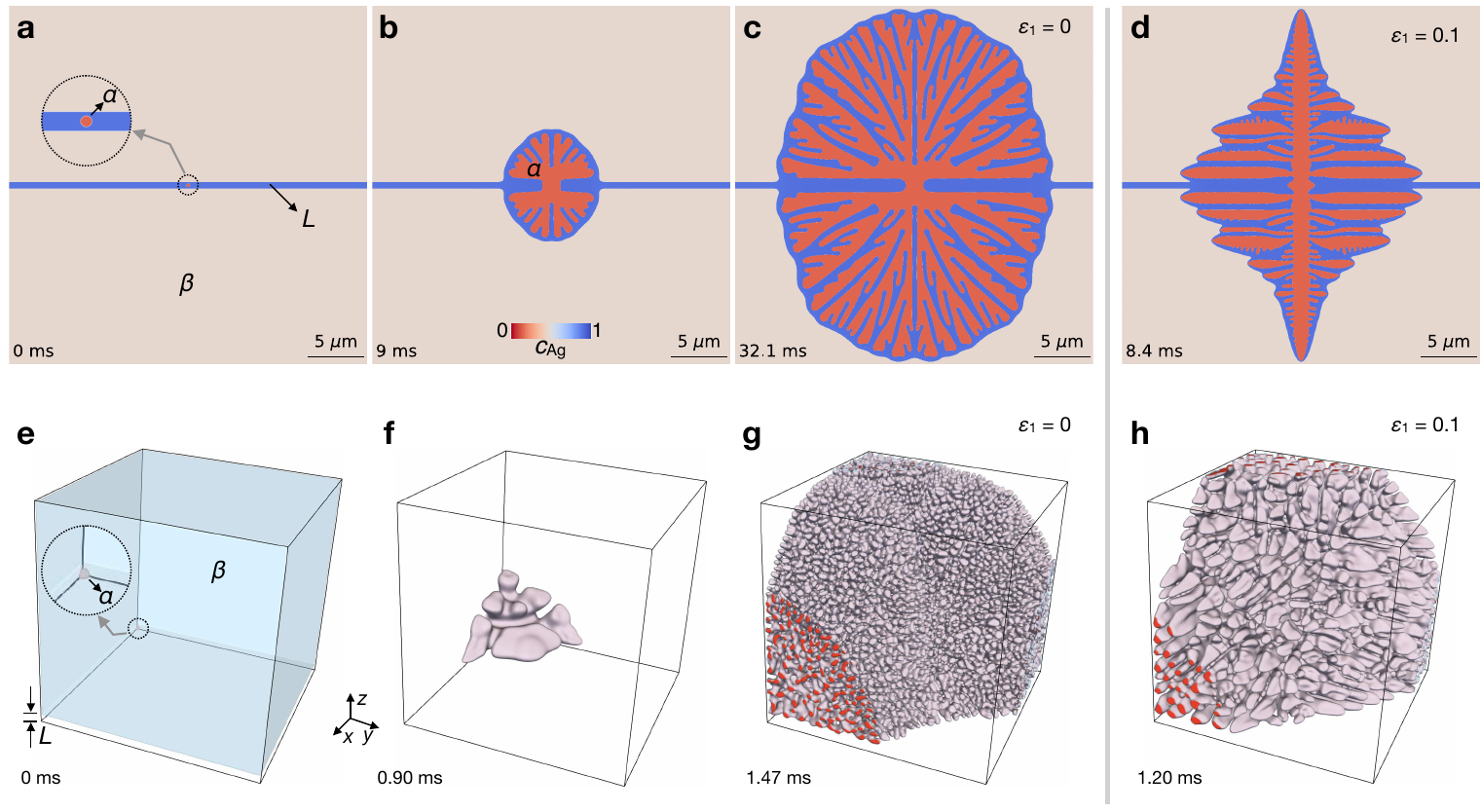}
\caption{\textbf{Liquid film migration instability during Ti--Ag peritectic melting.}
A Ti-rich $\alpha$ nucleus is placed in an Ag-rich liquid film between parent $\beta$ grains at  $\Delta T=100$ K above the peritectic temperature $T_p$. \textbf{(a--c)} In 2D with isotropic $\alpha$--liquid interfacial energy, the advancing interface destabilizes into a seaweed morphology. \textbf{(d)} Fourfold anisotropy selects dendritic branches. \textbf{(e--g)} In 3D, side-branch coalescence converts the unstable front into a bicontinuous $\alpha/L$ network. \textbf{(h)} Fourfold anisotropy biases growth along cubic axes while preserving bicontinuity.}
\label{fig:1}
\end{figure*}

\section*{Results and Discussion}

\subsection*{Morphological instability of liquid film migration and bicontinuous structure formation}

As observed experimentally, melting preceding $\alpha$ nucleation is typically initiated at grain boundaries \cite{hu2019evolution, li2025peritectic}, consistent with the theoretical expectation that the formation of a liquid film becomes energetically favored at a grain boundary when its excess free-energy $\gamma_{gb}> 2\gamma_{l\beta}$. For this reason, melting is only initiated at a subset of grain boundaries of the initial polycrystalline $\beta$ microstructure. Since we are interested in elucidating the topology-generating pattern formation mechanism associated with LFM, we do not model the details of the liquid film formation and $\alpha$ nucleation process. We simply start the simulations from a small initially 2D-circular or 3D-spherical nucleus immersed in a straight liquid film between two $\beta$ grains. The evolution of such a seed nucleus is shown in Figure~\ref{fig:1} in 2D (a-d) and 3D (e-h).  
In both 2D and 3D, the advancing $\alpha$--liquid interface destabilizes while remaining separated from the parent $\beta$ alloy by a continuous liquid film, forming a branched seaweed when $\gamma_{l\alpha}$ is isotropic and a dendritic morphology when $\gamma_{l\alpha}$ is anisotropic.

\begin{figure*}[!ht]
\centering
\includegraphics[width=0.7\textwidth]{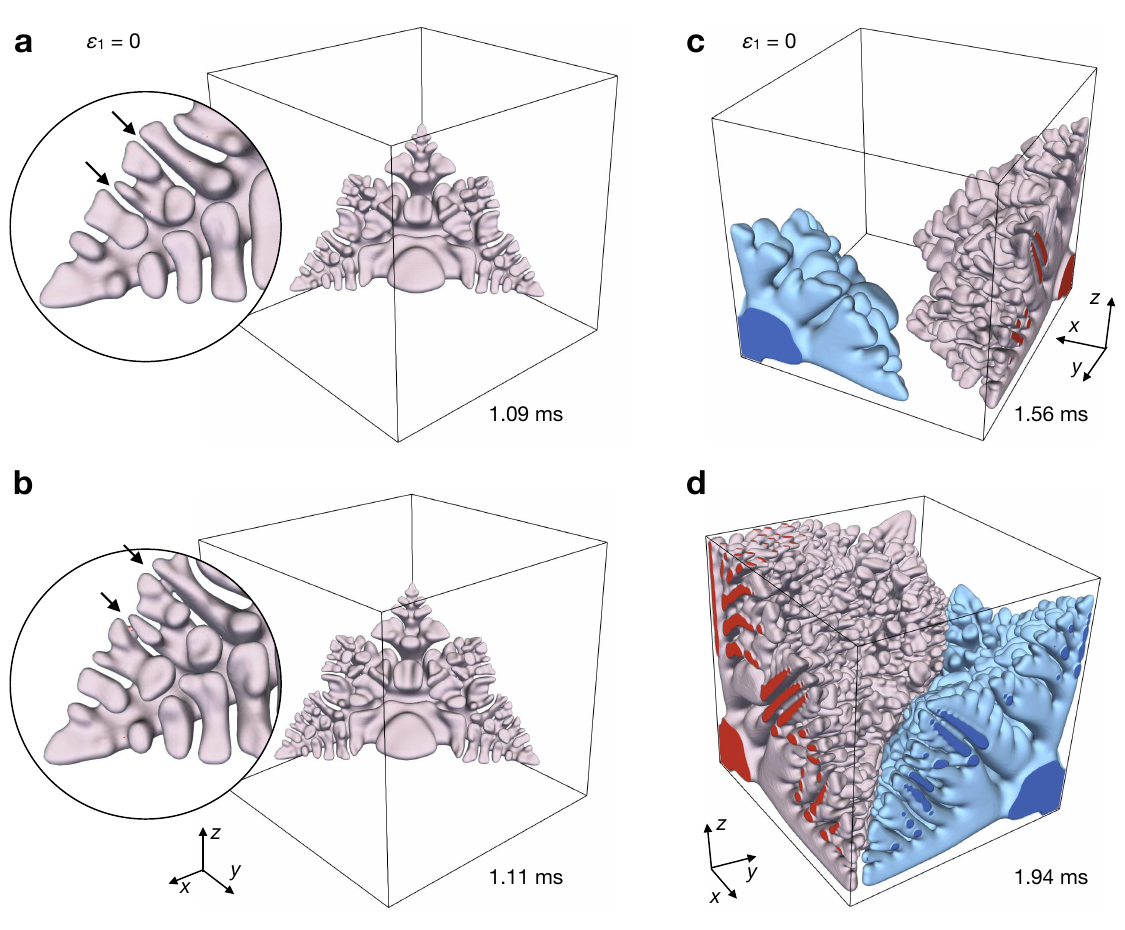}
\caption{\textbf{Bicontinuous network formation by side-branch coalescence.}
3D phase-field simulations at $\Delta T=100$ K. \textbf{(a,b)} In a single-nucleus simulation, neighboring $\alpha$ side branches expand laterally and coalesce, closing loops and creating topological handles. \textbf{(c,d)} In a two-nucleus simulation, separate dendritic envelopes coalesce internally and then impinge, producing a polycrystalline bicontinuous network.}
\label{fig:2}
\end{figure*}

These distinct branched morphologies are qualitatively expected from previous studies of single-phase dendritic growth that have highlighted the critical role of crystalline anisotropy in enabling dendritic growth along preferred crystal axes \cite{barbieri1989predictions,haxhimali2006orientation}, and the formation of seaweed structures in the absence of anisotropy \cite{ihle1994fractal,akamatsu1995symmetry}.
However, this role is only secondary here to the formation mechanism of bicontinuous structures, which can be generated in 3D by both seaweed and dendritic growth as illustrated in Figure~\ref{fig:1}g,h. More important is the topological transformation mediated by the coalescence of neighboring branches that is further illustrated in Figure~\ref{fig:2}a,b. On the time scale of the simulations of Figure~\ref{fig:1}d and h, coalescence is only present in the 3D geometry of experimental relevance. This difference can be attributed to the existence of two principal radii of curvature of the $\alpha$--liquid boundary in 3D, which facilitates transport of excess Ag in the liquid film, enabling $\alpha$--liquid interfaces to approach closely enough to coalesce in 3D but not in 2D.
Each coalescence event closes a loop in the $\alpha$ scaffold and creates a topological handle. Such side-branch coalescence is a known pathway for microstructural connectivity during classical dendritic solidification \cite{limodin2009situ}, but here it occurs within a migrating liquid film and generates bicontinuity during PM. Repetition of this elementary operation across the dealloying front converts a collection of dendritic branches into a connected bicontinuous $\alpha$/liquid network (Figure~\ref{fig:1}g). Anisotropy ($\epsilon_{1}=0.1$, Figure~\ref{fig:1}h) biases growth along the cubic axes but does not suppress branch coalescence, so that the $\alpha$ scaffold remains topologically interconnected. 

The same mechanism also explains how polycrystalline bicontinuous structures can form. When two $\alpha$ nuclei with distinct orientations are introduced in the same liquid film, each develops its own dendritic envelope. Branch coalescence first creates connectivity within each envelope, and subsequent impingement produces a polycrystalline bicontinuous network (Figure~\ref{fig:2}c,d). In this case, the grain scale is set primarily by the initial spacing of $\alpha$ nuclei, while the ligament scale is set by branching and coarsening of the migrating $\alpha$--liquid interface. PM therefore naturally separates grain-size selection from ligament-size selection, a distinction that may be important for designing polycrystalline bicontinuous microstructures.

\subsection*{Topology and coarsening}

\begin{figure*}[!hb]
\centering
\includegraphics[width=0.7\textwidth]{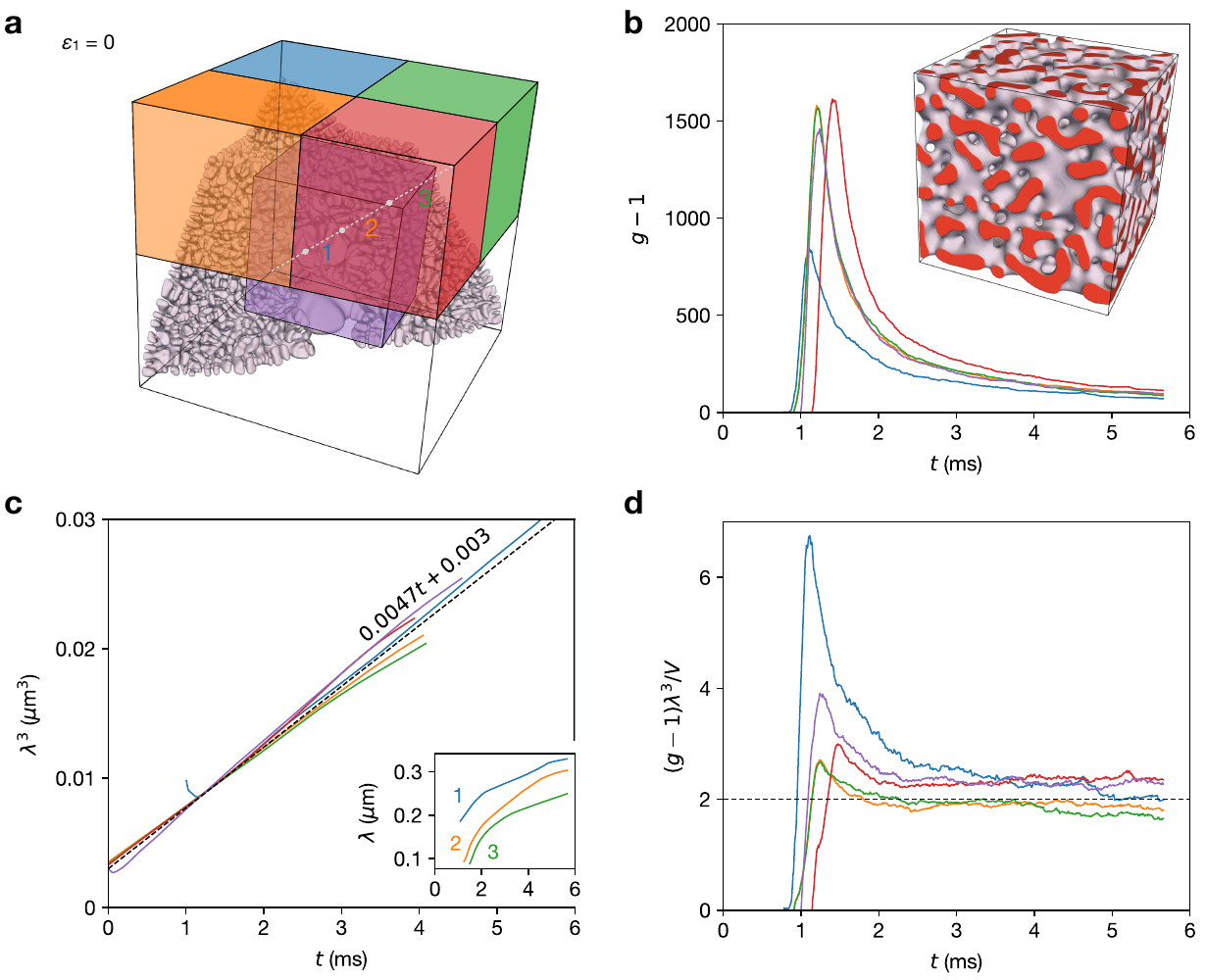}
\caption{\textbf{Topology and coarsening of the bicontinuous network.} \textbf{(a)} Sub-volumes used to measure the genus $g$ and ligament size $\lambda$.  \textbf{(b)} Evolution of $g-1$, confirming a high-genus topology; the inset shows the final morphology.  \textbf{(c)} Evolution of the cubed ligament size $\lambda^3$, with curves shifted in time to align local coarsening ages. The inset corresponds to the cube regions marked by diagonal segments 1--3 in panel (a).  \textbf{(d)} The scaled genus $(g-1)\lambda^3/V$, where $V$ is the analyzed volume, approaches $\sim 2$.}
\label{fig:3}
\end{figure*}

To confirm bicontinuity quantitatively, we computed the genus $g$ of the $\alpha$--liquid interface in cubic sub-volumes of the dealloyed region (Figure~\ref{fig:3}a). Genus counts independent handles of a closed surface: $g=0$ corresponds to a sphere-like topology, $g=1$ to a torus, and $g\gg1$ to a highly connected network. Details of the marching-cubes reconstruction and genus calculation are given in \suppnoteref{note:analysis}.

The genus increases rapidly during the active branching and coalescence stage, reaching values far above unity in all analyzed regions (Figure~\ref{fig:3}b). This confirms that the $\alpha$ scaffold is not merely branched, but topologically bicontinuous. At later times, $g$ decreases gradually as capillarity-driven coarsening removes small ligaments and eliminates some handles, while remaining well above unity. Thus, the network retains bicontinuity even as its characteristic length scale increases in time.

The ligament size $\lambda$, extracted from the structure factor of the $\alpha$--liquid interface (\suppnoteref{note:analysis}), follows the Lifshitz--Slyozov--Wagner cubic coarsening law \cite{lifshitz1961kinetics, wagner1961theory}
\begin{equation}
\lambda^{3} - \lambda(0)^{3} = k\,(t - t_{0}),
\label{eq:LSW}
\end{equation}
with a fitted coarsening constant $k=4.7~\mu\mathrm{m}^{3}/\mathrm{s}$ at superheating $\Delta T=100$ K (Figure~\ref{fig:3}c). This value predicts a ligament size of approximately $6.5~\mu$m after one minute of dealloying, consistent with experimental observations of Ti--Ag PM \cite{hu2019evolution, li2025peritectic}. Sub-volumes closer to the initial $\alpha$ nucleus have larger $\lambda$ at a given time because those ligaments entered the coarsening regime earlier (inset of Figure~\ref{fig:3}c), providing an internal age gradient within the same simulation.

Combining topology and length scale through the dimensionless quantity $(g-1)\lambda^{3}/V$, where $V$ is the analyzed volume, yields a value approaching $\sim 2$ at long times (Figure~\ref{fig:3}d). This near-constant scaled genus indicates self-similar coarsening of the bicontinuous structure and is consistent with previous PF studies of nanoporous-metal coarsening \cite{geslin2019phase}. The same behavior is obtained at $\Delta T=40$ K (Supplementary Figure~\ref{figs:40K}), where $g\gg1$, $\lambda^3$ grows linearly in time, and the scaled genus approaches a similar value, showing that bicontinuity and self-similar coarsening are robust consequences of the LFM instability.

\begin{figure*}[!hb]
\centering
\includegraphics[width=0.7\textwidth]{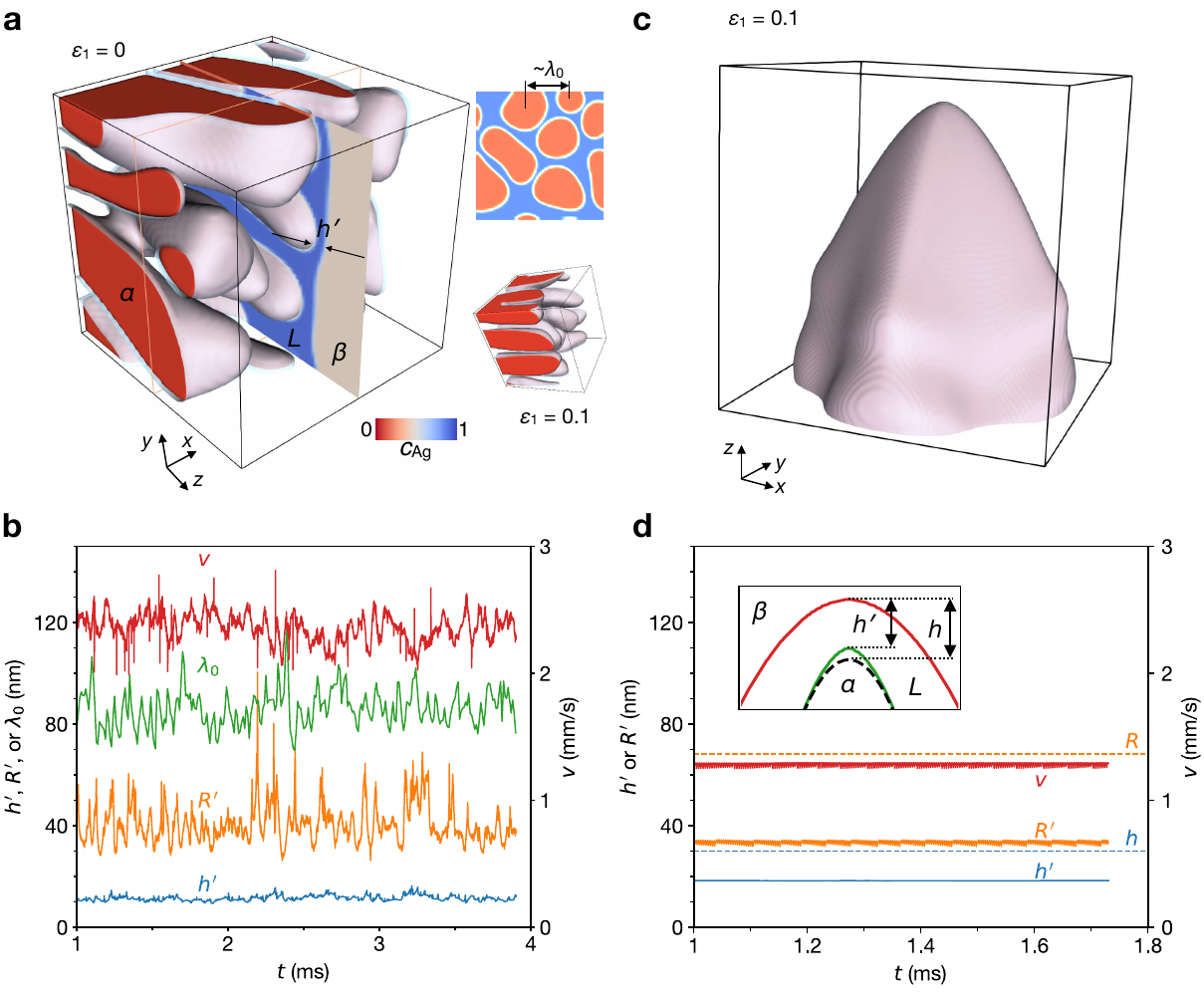}
\caption{\textbf{Tip kinetics during peritectic melting.}
\textbf{(a)} Moving-frame simulation of the multiple-dendrite LFM front. \textbf{(b)} The liquid-film thickness $h'$, tip radius $R'$, front-plane ligament size $\lambda_0$, and velocity $v$ remain nearly time-independent. \textbf{(c)} Single-dendrite LFM simulation. \textbf{(d)} In the single-dendrite case, $h'$ and $R'$ are measured directly from the interface, while $h$ and $R$ are obtained from a parabolic fit.}
\label{fig:4}
\end{figure*}

\subsection*{Tip kinetics}

A central distinction between PM and conventional LMD lies in the transport geometry. In LMD, the dissolved component must diffuse through the growing dealloyed layer to reach an external liquid medium, which leads to time-dependent kinetics, typically $v\propto1/\sqrt{t}$ \cite{geslin2015topology,lai2022topological}. In PM, no external reservoir exists. The Ag-rich liquid is generated locally at the melting front, and Ag rejected by the growing $\alpha$ phase is consumed locally by the receding $\beta$--liquid interface. This confines solute redistribution to the migrating liquid film. To investigate the resulting growth kinetics, we performed moving-frame simulations in which the computational domain follows the dealloying front (Figure~\ref{fig:4}a,b). This geometry allows the front velocity and length scales to be measured over long times. The multiple-dendrite LFM front contains many interacting $\alpha$ fingers, but the averaged quantities are well defined: the liquid-film thickness $h'$, tip radius $R'$, front-plane ligament size $\lambda_0$, and growth velocity $v$ all fluctuate around steady mean values. Their approximate time independence is the kinetic signature of local solute redistribution at the PM front and contrasts sharply with the diffusion-limited, time-dependent kinetics of LMD.

Although the multiple-dendrite front forms the bicontinuous network, it is too noisy for direct comparison with a tip-selection theory because individual fingers compete, branch, and coalesce. We therefore isolated the elementary growth unit by simulating a single anisotropy-selected $\alpha$ dendrite advancing through the same liquid-film geometry (Figure~\ref{fig:4}c,d and Supplementary Figure \ref{figs:3Dfit}). This reduced configuration assumes that single-dendrite growth kinetics is representative of the average behavior of the multiple-dendrite array, providing a steady LFM tip with well-defined velocity, liquid-film thickness, and curvature that can be directly compared with sharp-interface theory. The fact that isolated dendrites and multiple-dendrite arrays exhibit the same velocity-superheating relationship supports this assumption.

\begin{figure*}[!hb]
\centering
\includegraphics[width=\textwidth]{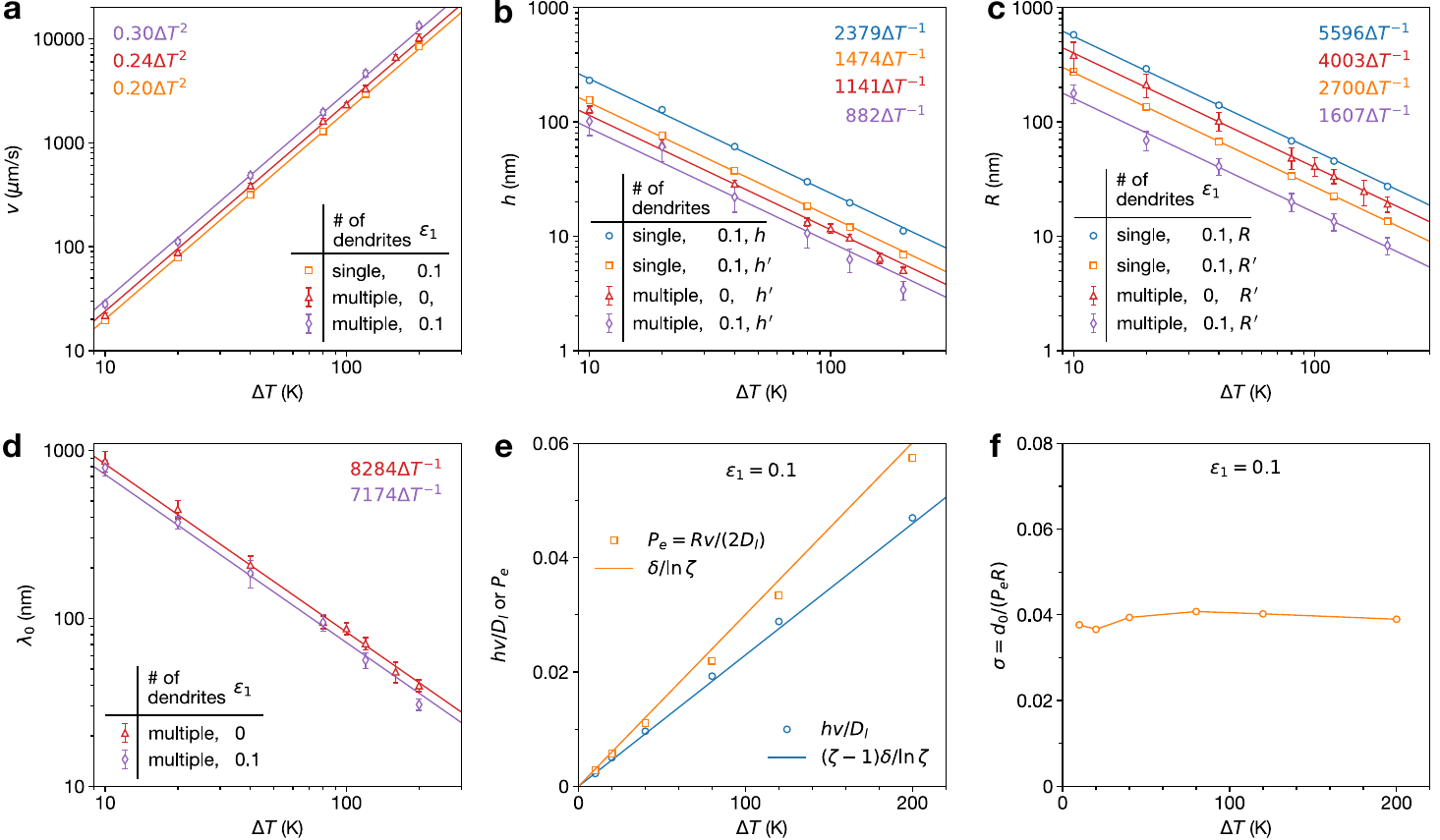}
\caption{\textbf{Scaling laws and solvability-type selection.} Symbols denote PF measurements for single and multiple dendrites; lines show fits or theoretical predictions. \textbf{(a)} $v\propto\Delta T^2$. \textbf{(b--d)} Film thickness, tip radius, and front-plane ligament size scale as $\Delta T^{-1}$. \textbf{(e)} $hv/D_l$ and $\mathrm{Pe}=Rv/(2D_l)$ agree with Eq.~(\ref{eq:hv_Pe}). \textbf{(f)} The stability parameter $\sigma=d_0/(\mathrm{Pe}R)$ is approximately constant.}
\label{fig:5}
\end{figure*}

The superheating dependence of these kinetic quantities is summarized in Figure~\ref{fig:5}. For both single- and multiple-dendrite simulations, the growth velocity follows $v\propto\Delta T^2$, while the liquid-film thickness, tip radius, and initial ligament size scale approximately as $\Delta T^{-1}$. The agreement of these scalings across  single and multiple (i.e. array of) primary dendrite branches indicates that the bicontinuous front inherits the same tip-scale selection physics branch by branch, even though branch interactions determine the final network topology.
To model the scalings obtained in our PF simulations, we extend the existing sharp-interface theory of dendritic growth \cite{ivantsov1947temperature,barbieri1989predictions} to LFM in 3D; previous studies only considered this extension in 2D \cite{brener2005velocity,brener2007melting,boussinot2010kinetics} (Supplementary Figure \ref{figs:2D3D}).
The theory combines two key ingredients: a solution of the diffusion equation 
in a frame translating at velocity $v$ with mass conservation conditions
at the moving $\alpha$-- and $\beta$--liquid interfaces assumed to have steady-state parabolic shapes in this frame, which is a direct extension of Ivantsov's transport theory for a single interface \cite{ivantsov1947temperature}, and a solvability condition that takes into account the effect of anisotropy \cite{barbieri1989predictions}.
The diffusion equation solution yields the relations 
\begin{equation}
\frac{h\,v}{D_{l}} = \frac{(\zeta - 1)\,\delta}{\ln\zeta},
\qquad
\mathrm{Pe} = \frac{\delta}{\ln\zeta},
\label{eq:hv_Pe}
\end{equation}
where $\mathrm{Pe}=vR/(2D_l)$ is the Peclet number, which is assumed small, $\mathrm{Pe}\ll 1$, 
$\delta=(c_{l\alpha}-c_{l\beta})/(c_p-c_{p\alpha})$ is the dimensionless driving force fixed by the superheating $\Delta T$, and $\zeta=(c_p-c_{p\alpha})/(c_p-c_{p\beta})$ is fixed by the Ti--Ag phase diagram. Since $\delta\sim \Delta T$ near $T_p$, Eq.~(\ref{eq:hv_Pe}) predicts $hv/D_l\propto\Delta T$ and $\mathrm{Pe}\propto\Delta T$, in agreement with the single-dendrite simulations (Figure~\ref{fig:5}e). The solvability condition, in turn, predicts that $\sigma=d_0/(\mathrm{Pe}R)$ is a constant determined by the magnitude $\epsilon_1$ of anisotropy, consistent with the nearly constant value of $\sigma\approx0.04$ extracted from 3D simulations (Figure~\ref{fig:5}f). These results yield
\begin{equation}
R \propto \frac{d_{0}}{\sigma\delta},
\qquad
v \propto \frac{D_{l}\sigma\delta^{2}}{d_{0}},
\label{eq:Rv_scaling}
\end{equation}
thereby explaining the observed $R\sim h\sim\Delta T^{-1}$ and $v\sim\Delta T^2$ scalings. Full derivations and definitions of $d_0$, $\delta$, and $\zeta$ are given in \suppnoteref{note:solvability}.

\subsection*{Mechanistic implications and outlook}

These results establish PM as a self-contained dealloying pathway that is mechanistically distinct from ECD, VPD, and especially LMD. Like LMD, PM involves two solid phases and a liquid metal and produces a high-genus solid/liquid morphology that coarsens self-similarly. Unlike LMD, however, PM requires no external dealloying bath: the liquid is generated internally by decomposition of the peritectic alloy, and Ag redistribution is confined to the migrating liquid film. This local transport explains why the PM front velocity and characteristic length scales remain approximately constant in time, whereas LMD exhibits time-dependent kinetics controlled by long-range diffusion through the dealloyed layer.

PM also provides a natural route to polycrystalline bicontinuous structures. Nucleation at parent $\beta$ grain boundaries controls the number, spacing, and orientation of $\alpha$
grains, whereas morphological instability of LFM, branch coalescence, and coarsening determine the ligament scale within each grain. This decoupling is absent in many LMD systems (e.g., TaTi dealloyed by a Cu melt \cite{geslin2015topology}) where the secondary phase inherits the crystallographic orientation of the precursor alloy (primary phase), and bicontinuous structure formation is mediated by diffusion-limited coupled growth of solid and liquid phases rather than LFM \cite{geslin2015topology}. In contrast, in LMD systems where the crystal structures of the primary and secondary phases differ, polycrystalline bicontinuous networks typically form as in (FeCr)$_{70}$Ni$_{30}$
dealloyed in liquid Mg \cite{mokhtari2021situ} and in SiC dealloyed in liquid Ge \cite{greenidge2020porous}. In PM, where the parent and product solids are always distinct phases, grain size and ligament size are determined by independent mechanisms, whereas how these scales are selected in the analogous LMD systems remains an open question.

Beyond Ti–Ag, unstable LFM together with interface coalescence may underlie bicontinuous structure formation in other systems with a dealloying medium. For example, the formation of a bicontinuous structure of graphite and a Si-rich phase observed in SiC dealloyed by liquid Ge has been interpreted within the LMD framework \cite{greenidge2020porous}, but the mechanism has not been explicitly modeled to date. The present results suggest that the formation mechanism may be similar to the one elucidated here since the primary role of Ge is to lower the peritectic transformation temperature of SiC-Ge into graphite and a binary liquid; isothermal melting of SiC (without Ge) would be expected to form bicontinuous structures via LFM by a similar mechanism as Ti–Ag, albeit at much higher temperature. In this scenario, the presence of a dealloying medium (liquid Ge) would alter the dealloying kinetics with $v\sim 1/\sqrt{t}$ instead of a constant velocity, but topology would be generated by unstable LFM together with interface coalescence instead of unstable two-phase coupled growth \cite{geslin2015topology}. The grain size, in turn, would be controlled by nucleation potency independently of the ligament size. The present results enlarge the scope of possible self-organized mechanisms of bicontinuous structure formation during dealloying. Which mechanism is at play in different multicomponent alloy systems remains to be investigated.

\section*{Conclusions}

Phase-field simulations demonstrate that isothermal melting of an initially homogeneous Ti–Ag peritectic alloy produces a high-genus bicontinuous structure through two sequential mechanisms: a morphological instability of liquid film migration that generates a highly branched structure of the Ti-rich secondary phase, followed by interface coalescence that transforms its topology. The simulations further reveal that 
the solidification front velocity and the initial ligament width at this front are constant in time, with subsequent coarsening following a $t^{1/3}$ law that predicts a final ligament width in quantitative agreement with experiment. The front velocity and initial ligament width follow scaling laws as functions of superheating in quantitative agreement with sharp-interface theory. These results establish PM as a distinct, internally driven dealloying pathway to bicontinuous structures, with grain size and ligament scale governed by independent mechanisms that can be controlled for microstructure design. More broadly, unstable LFM may underlie bicontinuous structure formation in other alloy systems even in the presence of a dealloying medium and decreasing velocity.

\section*{Acknowledgments}
This work was funded by the U.S. Department of Energy, Office of Science, Office of Basic Energy Sciences under Award Numbers DE-SC0020895. A.K. wishes to thank Zhongyang Li for valuable discussions. All numerical simulations were carried out on the Explorer cluster of Northeastern University at Massachusetts Green High Performance Computing Center (MGHPCC) in Holyoke, MA.

\section*{Declaration of Competing Interests}
The authors report there are no competing interests to declare.

\section*{Use of AI tools}

The authors used Claude Opus 4.7 (Anthropic) in two ways during preparation of this manuscript. First, for copyediting and language refinement of the main text and supplementary notes. Second, for coding assistance with the post-processing scripts used in the topology and structure-factor analyses. The tool was not used to generate scientific content, results, figures, data, or references. After using this tool, the authors reviewed and edited all output and take full responsibility for the integrity, accuracy, and originality of the content of this publication.

\bibliographystyle{unsrt}
\bibliography{reference}

@article{lifshitz1961kinetics,
  title={The kinetics of precipitation from supersaturated solid solutions},
  author={Lifshitz, Ilya M and Slyozov, Vitaly V},
  journal={Journal of physics and chemistry of solids},
  volume={19},
  number={1-2},
  pages={35--50},
  year={1961},
  publisher={Elsevier}
}

@article{hu2019evolution,
  title={Evolution of a bicontinuous structure in peritectic melting: The simplest form of dealloying},
  author={Hu, Wen-Kai and Shao, Jun-Chao and Wang, Shao-Gang and Jin, Hai-Jun},
  journal={Physical Review Materials},
  volume={3},
  number={11},
  pages={113601},
  year={2019},
  publisher={APS}
}

@article{geslin2015topology,
  title={Topology-generating interfacial pattern formation during liquid metal dealloying},
  author={Geslin, Pierre-Antoine and McCue, Ian and Gaskey, Bernard and Erlebacher, Jonah and Karma, Alain},
  journal={Nature communications},
  volume={6},
  number={1},
  pages={1--8},
  year={2015},
  publisher={Nature Publishing Group}
}

@article{geslin2019phase,
  title={Phase-field investigation of the coarsening of porous structures by surface diffusion},
  author={Geslin, Pierre-Antoine and Buchet, Mickael and Wada, Takeshi and Kato, Hidemi},
  journal={Physical Review Materials},
  volume={3},
  number={8},
  pages={083401},
  year={2019},
  publisher={APS}
}

@article{folch2005quantitative,
  title={Quantitative phase-field modeling of two-phase growth},
  author={Folch, R and Plapp, M},
  journal={Physical Review E},
  volume={72},
  number={1},
  pages={011602},
  year={2005},
  publisher={APS}
}

@article{dinsdale1991sgte,
  title={{SGTE} data for pure elements},
  author={Dinsdale, Alan T},
  journal={{CALPHAD}},
  volume={15},
  number={4},
  pages={317--425},
  year={1991},
  publisher={Elsevier}
}

@article{boussinot2010kinetics,
  title={Kinetics of isothermal phase transformations above and below the peritectic temperature: Phase-field simulations},
  author={Boussinot, G and Brener, EA and Temkin, DE},
  journal={Acta materialia},
  volume={58},
  number={5},
  pages={1750--1760},
  year={2010},
  publisher={Elsevier}
}

@inproceedings{ivantsov1947temperature,
  title={The temperature field around a spherical, cylindrical, or pointed crystal growing in a cooling solution},
  author={Ivantsov, GP},
  booktitle={Dokl. Akad. Nauk SSSR},
  volume={58},
  pages={567--569},
  year={1947}
}

@article{mccue2018pattern,
  title={Pattern formation during electrochemical and liquid metal dealloying},
  author={McCue, Ian and Karma, Alain and Erlebacher, Jonah},
  journal={Mrs Bulletin},
  volume={43},
  number={1},
  pages={27--34},
  year={2018},
  publisher={Cambridge University Press}
}

@article{wada2014bulk,
  title={Bulk-nanoporous-silicon negative electrode with extremely high cyclability for lithium-ion batteries prepared using a top-down process},
  author={Wada, Takeshi and Ichitsubo, Tetsu and Yubuta, Kunio and Segawa, Haruhiko and Yoshida, Hirokazu and Kato, Hidemi},
  journal={Nano letters},
  volume={14},
  number={8},
  pages={4505--4510},
  year={2014},
  publisher={ACS Publications}
}

@article{lu2018three,
  title={Three-dimensional bicontinuous nanoporous materials by vapor phase dealloying},
  author={Lu, Zhen and Li, Cheng and Han, Jiuhui and Zhang, Fan and Liu, Pan and Wang, Hao and Wang, Zhili and Cheng, Chun and Chen, Linghan and Hirata, Akihiko and others},
  journal={Nature communications},
  volume={9},
  number={1},
  pages={1--7},
  year={2018},
  publisher={Nature Publishing Group}
}

@article{shao2020nanoporous,
  title={Nanoporous carbon for electrochemical capacitive energy storage},
  author={Shao, Hui and Wu, Yih-Chyng and Lin, Zifeng and Taberna, Pierre-Louis and Simon, Patrice},
  journal={Chemical Society Reviews},
  volume={49},
  number={10},
  pages={3005--3039},
  year={2020},
  publisher={Royal Society of Chemistry}
}

@article{guo2008nanostructured,
  title={Nanostructured materials for electrochemical energy conversion and storage devices},
  author={Guo, Yu-Guo and Hu, Jin-Song and Wan, Li-Jun},
  journal={Advanced Materials},
  volume={20},
  number={15},
  pages={2878--2887},
  year={2008},
  publisher={Wiley Online Library}
}

@article{zielasek2006gold,
  title={Gold catalysts: nanoporous gold foams},
  author={Zielasek, Volkmar and J{\"u}rgens, Birte and Schulz, Christian and Biener, J{\"u}rgen and Biener, Monika M and Hamza, Alex V and B{\"a}umer, Marcus},
  journal={Angewandte Chemie International Edition},
  volume={45},
  number={48},
  pages={8241--8244},
  year={2006},
  publisher={Wiley Online Library}
}

@article{wittstock2010nanoporous,
  title={Nanoporous gold catalysts for selective gas-phase oxidative coupling of methanol at low temperature},
  author={Wittstock, A and Zielasek, V and Biener, J and Friend, CM and B{\"a}umer, M},
  journal={Science},
  volume={327},
  number={5963},
  pages={319--322},
  year={2010},
  publisher={American Association for the Advancement of Science}
}

@article{hodge2007scaling,
  title={Scaling equation for yield strength of nanoporous open-cell foams},
  author={Hodge, AM and Biener, J and Hayes, JR and Bythrow, PM and Volkert, CA and Hamza, AV},
  journal={Acta Materialia},
  volume={55},
  number={4},
  pages={1343--1349},
  year={2007},
  publisher={Elsevier}
}

@article{jin2009deforming,
  title={Deforming nanoporous metal: Role of lattice coherency},
  author={Jin, Hai-Jun and Kurmanaeva, Lilia and Schmauch, J{\"o}rg and R{\"o}sner, Harald and Ivanisenko, Yulia and Weissm{\"u}ller, J{\"o}rg},
  journal={Acta Materialia},
  volume={57},
  number={9},
  pages={2665--2672},
  year={2009},
  publisher={Elsevier}
}

@article{hu2008electrochemical,
  title={Electrochemical DNA biosensor based on nanoporous gold electrode and multifunctional encoded DNA- Au bio bar codes},
  author={Hu, Kongcheng and Lan, Dongxiao and Li, Xuemei and Zhang, Shusheng},
  journal={Analytical chemistry},
  volume={80},
  number={23},
  pages={9124--9130},
  year={2008},
  publisher={ACS Publications}
}

@article{zhang2012nanoporous,
  title={Nanoporous metals: fabrication strategies and advanced electrochemical applications in catalysis, sensing and energy systems},
  author={Zhang, Jintao and Li, Chang Ming},
  journal={Chemical Society Reviews},
  volume={41},
  number={21},
  pages={7016--7031},
  year={2012},
  publisher={Royal Society of Chemistry}
}

@article{erlebacher2001evolution,
  title={Evolution of nanoporosity in dealloying},
  author={Erlebacher, Jonah and Aziz, Michael J and Karma, Alain and Dimitrov, Nikolay and Sieradzki, Karl},
  journal={Nature},
  volume={410},
  number={6827},
  pages={450--453},
  year={2001},
  publisher={Nature Publishing Group}
}

@article{snyder2010oxygen,
  title={Oxygen reduction in nanoporous metal--ionic liquid composite electrocatalysts},
  author={Snyder, J and Fujita, T and Chen, MW and Erlebacher, J},
  journal={Nature materials},
  volume={9},
  number={11},
  pages={904--907},
  year={2010},
  publisher={Nature Publishing Group}
}

@article{hayes2006monolithic,
  title={Monolithic nanoporous copper by dealloying {Mn--Cu}},
  author={Hayes, JR and Hodge, AM and Biener, J and Hamza, AV and Sieradzki, Karl},
  journal={Journal of Materials Research},
  volume={21},
  number={10},
  pages={2611--2616},
  year={2006},
  publisher={Cambridge University Press}
}

@article{wada2011dealloying,
  title={Dealloying by metallic melt},
  author={Wada, Takeshi and Yubuta, Kunio and Inoue, Akihisa and Kato, Hidemi},
  journal={Materials Letters},
  volume={65},
  number={7},
  pages={1076--1078},
  year={2011},
  publisher={Elsevier}
}

@article{mccue2016size,
  title={Size effects in the mechanical properties of bulk bicontinuous {Ta/Cu} nanocomposites made by liquid metal dealloying},
  author={McCue, Ian and Ryan, Stephen and Hemker, Kevin and Xu, Xiandong and Li, Nan and Chen, Mingwei and Erlebacher, Jonah},
  journal={Advanced Engineering Materials},
  volume={18},
  number={1},
  pages={46--50},
  year={2016},
  publisher={Wiley Online Library}
}

@article{lu2021vapor,
  title={Vapor phase dealloying kinetics of MnZn alloys},
  author={Lu, Zhen and Zhang, Fan and Wei, Daixiu and Han, Jiuhui and Xia, Yanjie and Jiang, Jing and Zhong, Mingwang and Hirata, Akihiko and Watanabe, Kentaro and Karma, Alain and others},
  journal={Acta Materialia},
  volume={212},
  pages={116916},
  year={2021},
  publisher={Elsevier}
}

@article{han2019vapor,
  title={Vapor phase dealloying: A versatile approach for fabricating 3D porous materials},
  author={Han, Jiuhui and Li, Cheng and Lu, Zhen and Wang, Hao and Wang, Zhili and Watanabe, Kentaro and Chen, Mingwei},
  journal={Acta Materialia},
  volume={163},
  pages={161--172},
  year={2019},
  publisher={Elsevier}
}

@article{li2005experimental,
  title={Experimental study and thermodynamic assessment of the {Ag--Ti} system},
  author={Li, Mei and Li, Changrong and Wang, Fuming and Zhang, Weijing},
  journal={ {CALPHAD} },
  volume={29},
  number={4},
  pages={269--275},
  year={2005},
  publisher={Elsevier}
}

@article{lai2022topological,
  title={Topological control of liquid-metal-dealloyed structures},
  author={Lai, Longhai and Gaskey, Bernard and Chuang, Alyssa and Erlebacher, Jonah and Karma, Alain},
  journal={Nature Communications},
  volume={13},
  number={1},
  pages={2918},
  year={2022},
  publisher={Nature Publishing Group UK London}
}

@article{haxhimali2006orientation,
  title={Orientation selection in dendritic evolution},
  author={Haxhimali, Tomorr and Karma, Alain and Gonzales, Fr{\'e}d{\'e}ric and Rappaz, Michel},
  journal={Nature materials},
  volume={5},
  number={8},
  pages={660--664},
  year={2006},
  publisher={Nature Publishing Group}
}

@article{lai2022microstructural,
  title={Microstructural pattern formation during liquid metal dealloying: Phase-field simulations and theoretical analyses},
  author={Lai, Longhai and Geslin, Pierre-Antoine and Karma, Alain},
  journal={Physical Review Materials},
  volume={6},
  number={9},
  pages={093803},
  year={2022},
  publisher={APS}
}

@article{harrison1959attack,
  title={The attack of solid alloys by liquid metals and salt melts},
  author={Harrison, John David and Wagner, C},
  journal={Acta Metallurgica},
  volume={7},
  number={11},
  pages={722--735},
  year={1959},
  publisher={Elsevier}
}

@article{kim2015optimizing,
  title={Optimizing niobium dealloying with metallic melt to fabricate porous structure for electrolytic capacitors},
  author={Kim, Joung Wook and Tsuda, Masashi and Wada, Takeshi and Yubuta, Kunio and Kim, Sung Gyoo and Kato, Hidemi},
  journal={Acta Materialia},
  volume={84},
  pages={497--505},
  year={2015},
  publisher={Elsevier}
}

@article{greenidge2020porous,
  title={Porous graphite fabricated by liquid metal dealloying of silicon carbide},
  author={Greenidge, G and Erlebacher, J},
  journal={Carbon},
  volume={165},
  pages={45--54},
  year={2020},
  publisher={Elsevier}
}

@article{muschik1989melting,
  title={Melting of {Cu--In} solid solutions at small superheating by droplet formation and liquid film migration},
  author={Muschik, T and Kaysser, WA and Hehenkamp, T},
  journal={Acta Metallurgica},
  volume={37},
  number={2},
  pages={603--613},
  year={1989},
  publisher={Elsevier}
}

@article{mokhtari2021situ,
  title={In situ observation of liquid metal dealloying and etching of porous FeCr by X-ray tomography and X-ray diffraction},
  author={Mokhtari, Morgane and Le Bourlot, Christophe and Adrien, J{\'e}rome and Bonnin, Anne and Ludwig, Wolfgang and Geslin, Pierre-Antoine and Wada, Takeshi and Duchet-Rumeau, Jannick and Kato, Hidemi and Maire, Eric},
  journal={Materialia},
  volume={18},
  pages={101125},
  year={2021},
  publisher={Elsevier}
}

@incollection{lorensen1998marching,
  title={Marching cubes: A high resolution 3D surface construction algorithm},
  author={Lorensen, William E and Cline, Harvey E},
  booktitle={Seminal graphics: pioneering efforts that shaped the field},
  pages={347--353},
  year={1998}
}

@article{brener2005velocity,
  title={Velocity-selection problem for combined motion of melting and solidification fronts},
  author={Brener, Efim A and Temkin, DE},
  journal={Physical review letters},
  volume={94},
  number={18},
  pages={184501},
  year={2005},
  publisher={APS}
}

@article{brener2007melting,
  title={Melting of alloys along the inter-phase boundaries in eutectic and peritectic systems},
  author={Brener, Efim A and Temkin, DE},
  journal={Acta materialia},
  volume={55},
  number={8},
  pages={2785--2789},
  year={2007},
  publisher={Elsevier}
}

@article{song2022ultrafine,
  title={Ultrafine nanoporous intermetallic catalysts by high-temperature liquid metal dealloying for electrochemical hydrogen production},
  author={Song, Ruirui and Han, Jiuhui and Okugawa, Masayuki and Belosludov, Rodion and Wada, Takeshi and Jiang, Jing and Wei, Daixiu and Kudo, Akira and Tian, Yuan and Chen, Mingwei and others},
  journal={Nature Communications},
  volume={13},
  number={1},
  pages={5157},
  year={2022},
  publisher={Nature Publishing Group UK London}
}

@article{li2025peritectic,
  title={How peritectic melting forms bicontinuous microstructures},
  author={Li, Zhongyang and L{\"u}hrs, Lukas and Krekeler, Tobias and Weissm{\"u}ller, J{\"o}rg},
  journal={Acta Materialia},
  volume={289},
  pages={120917},
  year={2025},
  publisher={Elsevier}
}

@article{zhong2025quantification,
  title={Quantification and prediction of solidification textures under additive manufacturing conditions},
  author={Zhong, Mingwang and Eres-Castellanos, Adriana and Ji, Kaihua and Saville, Alec I and Rodgers, Brian and Coughlin, Dan R and Gibbs, John W and Roehling, John D and McKeown, Joseph T and Clarke, Amy J and others},
  journal={Nature Communications},
  year={2025},
  publisher={Nature Publishing Group UK London}
}

@article{barbieri1989predictions,
  title={Predictions of dendritic growth rates in the linearized solvability theory},
  author={Barbieri, A and Langer, JS},
  journal={Physical Review A},
  volume={39},
  number={10},
  pages={5314},
  year={1989},
  publisher={APS}
}

@article{tourret2015growth,
  title={Growth competition of columnar dendritic grains: A phase-field study},
  author={Tourret, Damien and Karma, Alain},
  journal={Acta Materialia},
  volume={82},
  pages={64--83},
  year={2015},
  publisher={Elsevier}
}

@article{ihle1994fractal,
  title={Fractal and compact growth morphologies in phase transitions with diffusion transport},
  author={Ihle, T and M{\"u}ller-Krumbhaar, H},
  journal={Physical Review E},
  volume={49},
  number={4},
  pages={2972},
  year={1994},
  publisher={APS}
}

@article{akamatsu1995symmetry,
  title={Symmetry-broken double fingers and seaweed patterns in thin-film directional solidification of a nonfaceted cubic crystal},
  author={Akamatsu, Silv{\`e}re and Faivre, Gabriel and Ihle, Thomas},
  journal={Physical Review E},
  volume={51},
  number={5},
  pages={4751},
  year={1995},
  publisher={APS}
}

@article{wagner1961theory,
  title={Theory of the aging of precipitates by dissolution-reprecipitation (Ostwald ripening)},
  author={Wagner, C},
  journal={Z Elektrochem},
  volume={65},
  number={7},
  pages={581--11},
  year={1961}
}

@article{limodin2009situ,
  title={In situ and real-time 3-D microtomography investigation of dendritic solidification in an Al--10 wt.\% Cu alloy},
  author={Limodin, Nathalie and Salvo, Luc and Boller, Elodie and Su{\'e}ry, Michel and Felberbaum, M and Gailli{\`e}gue, Sylvain and Madi, Kamel},
  journal={Acta Materialia},
  volume={57},
  number={7},
  pages={2300--2310},
  year={2009},
  publisher={Elsevier}
}

\clearpage



\renewcommand{\figurename}{Supplementary Figure} 
\renewcommand{\thefigure}{\arabic{figure}}
\renewcommand{\tablename}{Supplementary Table} 
\renewcommand{\thetable}{\arabic{table}}
\renewcommand{\theequation}{S\arabic{equation}}
\setcounter{figure}{0}
\setcounter{table}{0}
\setcounter{equation}{0}
\setcounter{page}{1}

\title{\textbf{Supplemental Materials: \\ Dealloying by peritectic melting}}
\maketitle
\tableofcontents


\supplementarynote{Phase-field modeling of peritectic melting}{note:PF}

We use the multi-phase-field model of Folch and Plapp~\cite{folch2005quantitative} adapted to the peritectic decomposition of Ti$_{50}$Ag$_{50}$, in which the intermetallic $\beta$ phase melts above $T_p$ into a Ti-rich solid $\alpha$ and an Ag-rich liquid. The state of the system is described by three non-conserved order parameters $\phi_\alpha$, $\phi_\beta$, and $\phi_l$ representing the local volume fractions of the three phases under the constraint $\phi_\alpha+\phi_\beta+\phi_l=1$, together with the conserved Ag concentration $c$. We adopt the index convention $\{\alpha,\beta,l\}\to\{1,2,3\}$ throughout. The free-energy functional is
\begin{equation}
\mathcal{F}=\int_{V}\!\left[\sum_{i}\frac{\sigma}{2}|\nabla\phi_{i}|^{2}+H\,f_{\phi}(\vec{\phi})+X\,f_{c}(c,\vec{\phi})\right]dV,
\label{eq:freeEnergy}
\end{equation}
where $\sigma$ is the gradient-energy coefficient that fixes the diffuse-interface width $W=\sqrt{\sigma/H}$, and $H$ and $X$ are energy-density scales. The multiwell potential
\begin{equation}
f_{\phi}=\sum_{i}\phi_{i}^{2}(1-\phi_{i})^{2}+\sum_{i}a_{i}\,\phi_{j}^{2}\phi_{k}^{2}\bigl(2\phi_{j}\phi_{k}+3\phi_{i}+b\,\phi_{i}^{2}\bigr)
\label{eq:fphi}
\end{equation}
has minima at the three bulk phases. The coefficient $a_{i}$ controls the interfacial free energy of the $jk$ interface that does not involve phase $i$, and $b$ raises the energy barrier at the triple junction, suppressing the spurious emergence of the third phase at any two-phase interface. The chemical free-energy density couples $\vec\phi$ to $c$ through
\begin{equation}
f_{c}=\frac{1}{2}\Bigl[c-\sum_{i}\hat{A}_{i}(T)\,g_{i}(\vec{\phi})\Bigr]^{2}+\sum_{i}\hat{B}_{i}(T)\,g_{i}(\vec{\phi}),
\label{eq:fc}
\end{equation}
with
\begin{equation}
g_{i}(\vec{\phi})=\frac{\phi_{i}}{4}\bigl\{15(1-\phi_{i})\bigl[1+\phi_{i}-(\phi_{j}-\phi_{k})^{2}\bigr]+\phi_{i}(9\phi_{i}^{2}-5)\bigr\}
\label{eq:gi}
\end{equation}
the lowest-order interpolation function that is antisymmetric about $\phi_{i}=1/2$ and satisfies $g_{i}(\phi_{i}=0)=0$, $g_{i}(\phi_{i}=1)=1$. The temperature-dependent coefficients $\hat{A}_{i}(T)$ and $\hat{B}_{i}(T)$ are chosen so that Eq.~(\ref{eq:fc}) reproduces the linearized Ti--Ag phase diagram near $T_{p}$:
\begin{equation}
\begin{aligned}
\hat{A}_{l}&=A_{1}, & \hat{B}_{l}&=B_{1}(T),\\
\hat{A}_{\alpha}&=-A_{1}-A_{2},\quad & \hat{B}_{\alpha}&=-B_{1}(T)-B_{2}(T),\\
\hat{A}_{\beta}&=-A_{1}+A_{2},\quad & \hat{B}_{\beta}&=-B_{1}(T)+B_{2}(T),
\end{aligned}
\end{equation}
with
\begin{equation}
\begin{aligned}
A_{1}&=\frac{1}{4}(1+r), & A_{2}&=\frac{1}{2}(1-r),\\
B_{1}&=B_{11}+B_{12}\widetilde{T}, & B_{2}&=B_{21}+B_{22}\widetilde{T}.
\end{aligned}
\end{equation}
Here $r=\widetilde{c}_{p}=(c-c_{p\beta})/(c_{p}-c_{p\alpha})|_{c=c_{p}}$ is the scaled liquidus concentration at $T_{p}$ and $\widetilde{T}=(T-T_{p})/[|m_{\alpha}|(c_{p}-c_{p\alpha})]$ is the scaled temperature. The phase-diagram parameters $T_{p}$, $c_{p\alpha}$, $c_{p\beta}$, $c_{p}$, $m_{\alpha}$, and $m_{\beta}$ are listed in Supplementary Table~\ref{tabs:constant} and visualized in Supplementary Figure~\ref{figs:phase_diagram}.

The conserved field evolves by a Cahn--Hilliard equation,
\begin{equation}
\frac{\partial c}{\partial t}=\nabla\!\cdot\!\bigl[M(\vec\phi)\,\nabla\mu\bigr],
\qquad
M(\vec\phi)=\frac{D_{l}\,\phi_{l}}{H},
\qquad
\mu=\frac{\delta\mathcal{F}}{\delta c},
\label{eq:c_evolution}
\end{equation}
in which the atomic mobility vanishes linearly with $\phi_{l}$ at the solid--liquid interfaces. The non-conserved phase fields follow a constrained Allen--Cahn equation,
\begin{equation}
\tau(\vec\phi)\,\frac{\partial\phi_{i}}{\partial t}
=-\frac{1}{H}\!\left(\frac{\delta\mathcal{F}}{\delta\phi_{i}}
-\frac{1}{3}\sum_{j}\frac{\delta\mathcal{F}}{\delta\phi_{j}}\right),
\label{eq:phi_evolution}
\end{equation}
in which the second term enforces $\sum_{i}\phi_{i}=1$ and the position-dependent time scale $\tau(\vec\phi)$ is chosen to eliminate the residual interface-thickness dependence of the simulated kinetics. Setting the effective interface kinetic coefficient at the $\alpha$--liquid and $\beta$--liquid interfaces to zero in the thin-interface limit (Eq.~(3.41) of Folch and Plapp~\cite{folch2005quantitative}) fixes two interface-specific time scales,
\begin{equation}
\tau_{\alpha}=0.7464\,\widetilde{\lambda}\,|\hat{A}_{l}-\hat{A}_{\alpha}|^{2}\,\frac{W^{2}}{D_{l}},
\qquad
\tau_{\beta}=0.7464\,\widetilde{\lambda}\,|\hat{A}_{l}-\hat{A}_{\beta}|^{2}\,\frac{W^{2}}{D_{l}},
\label{eq:tau_alpha_beta}
\end{equation}
where the numerical prefactor $0.7464$ comes from the matched asymptotic analysis of Folch and Plapp~\cite{folch2005quantitative}, and we define the average $\bar{\tau}=(\tau_{\alpha}+\tau_{\beta})/2$. To ensure that $\tau(\vec\phi)$ reduces to $\tau_{\alpha}$ inside the $\alpha$ bulk, $\tau_{\beta}$ inside the $\beta$ bulk, and $\bar{\tau}$ inside the liquid, we follow Eq.~(3.46) of Folch and Plapp~\cite{folch2005quantitative}:
\begin{equation}
\tau(\vec{\phi})=
\begin{cases}
\bar{\tau}+\dfrac{(\tau_{\beta}-\tau_{\alpha})(\phi_{\beta}-\phi_{\alpha})}{2(\phi_{\alpha}+\phi_{\beta})}, & \text{if }\phi_{l}\neq 1,\\[8pt]
\bar{\tau}, & \text{if }\phi_{l}=1.
\end{cases}
\label{eq:tau_interp}
\end{equation}

Equations~(\ref{eq:c_evolution}) and~(\ref{eq:phi_evolution}) are non-dimensionalized using $W$ as the length scale and $\bar{\tau}$ as the time scale, giving $\widetilde{x}=x/W$, $\widetilde{t}=t/\bar{\tau}$, $\widetilde\nabla=W\nabla$, $\widetilde\mu=\mu/H$, $\widetilde{\mathcal F}=\mathcal F/H$, and the dimensionless chemical prefactor $\widetilde{\lambda}=X/H$. The dimensionless equations are integrated on a uniform Cartesian grid with second-order central finite differences and explicit Euler time stepping. The time step is bounded by the diffusion stability limit,
\begin{equation}
d\widetilde{t}=\frac{0.6\,d\widetilde{x}^{2}}{2\,d_{\text{dim}}\,\widetilde{D}_{l}},
\qquad
\widetilde{D}_{l}\equiv\frac{D_{l}\,\bar{\tau}}{W^{2}},
\label{eq:dt}
\end{equation}
where $d_{\text{dim}}=2$ in 2D and $d_{\text{dim}}=3$ in 3D arises from the central-difference Laplacian of the explicit scheme, and the prefactor $0.6$ is chosen small enough to satisfy stability and large enough to accelerate the simulations. The code is implemented in CUDA C/C++; simulations are carried out on NVIDIA V100 or H200 GPUs.

\begin{table}[h]
\captionsetup{justification=centering}
\caption{\textbf{Parameters of the phase-field model}}
\makebox[\textwidth][c]{
\begin{tabular}{{cll}} 
\toprule
{Parameter} & {Description} & {Value}  \\ 
\midrule
$T_p$ & Peritectic temperature of the Ti-Ag system & 1293 K\\
$c_{p\alpha}$ & Ag composition of the $\alpha$ solidus at peritectic temperature  &  16.41 at.\% \\
$c_{p\beta}$ & Ag composition of the $\beta$ solid at peritectic temperature  & 50 at.\% \\
$c_p$ &  Ag composition of the liquidus at peritectic temperature  &  93.95 at.\%\\
$m_\alpha$ & Liquidus slope above peritectic temperature  & -40.078 K/at.\%\\
$m_\beta$ & Liquidus slope below peritectic temperature  & -26.187 K/at.\%  \\
$D_l$ & Diffusion constant in the liquid & $2\times 10^{-9}$ m\textsuperscript{2}/s \\
$\gamma_{\alpha l}$ & Interfacial free-energy for the liquid-$\alpha$ interface &0.20 J/m\textsuperscript{2}  \\
$\gamma_{\beta l}$ & Interfacial free-energy for the liquid-$\beta$ interface & 0.19 J/m\textsuperscript{2}  \\
$\gamma_{\alpha\beta}$ &Interfacial free-energy for the $\alpha$-$\beta$ interface & 0.18 J/m\textsuperscript{2}  \\
\bottomrule
\end{tabular}}
\label{tabs:constant}
\end{table}

\subsection*{Estimation of model parameters}

\paragraph{Interfacial free energies, $a_{i}$, and $b$.}
For each interface, the dimensionless surface free energy is set by the multiwell coefficient associated with the \emph{third} phase. With $\{\alpha,\beta,l\}\to\{1,2,3\}$, $\gamma_{\beta l}$ is controlled by $a_{1}$, $\gamma_{\alpha l}$ by $a_{2}$, and $\gamma_{\alpha\beta}$ by $a_{3}$, through~\cite{folch2005quantitative}
\begin{equation}
\widetilde{\gamma}_{jk}(a_{i})=\frac{2\sqrt{a_{i}}\,(3a_{i}+4)+(a_{i}+4)(3a_{i}-4)\,\mathrm{arccot}\!\bigl(2/\!\sqrt{a_{i}}\bigr)}{16\sqrt{2}\,a_{i}^{3/2}},
\qquad
\gamma_{jk}=\widetilde{\gamma}_{jk}\,W\!H.
\label{eq:ai}
\end{equation}
We use a single set of interfacial free energies throughout, $\gamma_{\beta l}=0.19$, $\gamma_{\alpha l}=0.20$, and $\gamma_{\alpha\beta}=0.18$~J/m$^{2}$, corresponding to $a_{1}=4.54285$, $a_{2}=5.58545$, and $a_{3}=3.55521$, respectively. The triple-junction coefficient is fixed at $b=80$, large enough to fully suppress the spurious appearance of any third phase at a two-phase interface.

\paragraph{Energy-density scale and $\widetilde{\lambda}$.}
The dimensionless prefactor $\widetilde{\lambda}=X/H$ is introduced as the relative weight of chemical and gradient-energy contributions in Eq.~(\ref{eq:freeEnergy}). Equivalently, in the scaled variables $\widetilde{c}=(c-c_{p\beta})/(c_{p}-c_{p\alpha})$ and $\widetilde{\mu}=\mu/H$, it is the slope of the equilibrium chemical potential with respect to concentration,
\begin{equation}
\widetilde{\lambda}=\left.\frac{\partial\widetilde{\mu}}{\partial\widetilde{c}}\right|_\text{eq}.
\label{eq:lambdaTilde_def}
\end{equation}
We calibrate $\widetilde{\lambda}$ by matching the latent heat of melting across the $\alpha$--liquid and $\beta$--liquid interfaces at $T_{p}$, which we compute in two independent ways and equate.

\emph{Route 1 — thermodynamic functions.} Equilibrium of phase $i$ and liquid at $T_{p}$ requires equal molar free energies,
\begin{equation}
G_{l}^{m}-G_{i}^{m}=0=H_{l}^{m}-H_{i}^{m}-T_{p}\bigl(S_{l}^{m}-S_{i}^{m}\bigr)=L_{i}^{m}-T_{p}\,\Delta S_{i}^{m},
\end{equation}
where $L_{i}^{m}=H_{l}^{m}-H_{i}^{m}$ is the latent heat of melting of phase $i$. Using the Gibbs--Helmholtz identity $S^{m}=-\partial G^{m}/\partial T$,
\begin{equation}
L_{i}^{m}=T_{p}\bigl(S_{l}^{m}-S_{i}^{m}\bigr)=T_{p}\!\left(\left.\frac{\partial G_{l}^{m}}{\partial T}\right|_{p}-\left.\frac{\partial G_{i}^{m}}{\partial T}\right|_{p}\right),
\label{eq:Li_thermo}
\end{equation}
which, evaluated with the published temperature-dependent free-energy functions $G_{l}^{m}(T)$, $G_{\alpha}^{m}(T)$, $G_{\beta}^{m}(T)$ of the Ti--Ag binary~\cite{dinsdale1991sgte,li2005experimental}, yields $L_{\alpha}=15.9$~kJ/mol and $L_{\beta}=28.3$~kJ/mol.

\emph{Route 2 — phase-field model.} The same latent heat can be obtained from the grand-potential balance across the $i$--liquid interface implied by Eqs.~(\ref{eq:freeEnergy})--(\ref{eq:fc}). Defining the molar volume $V_{m}$ and the equilibrium concentration jump $\Delta c_{\alpha}=c_{p}-c_{p\alpha}$, the molar free-energy difference takes the compact form
\begin{equation}
G_{l}^{m}-G_{i}^{m}=\widetilde{\mu}^{li}\,\Delta c_{\alpha}\,\bigl(\widetilde{c}_{l}^{li}-\widetilde{c}_{i}^{li}\bigr)\,H\,V_{m},
\label{eq:GLM_PF}
\end{equation}
where $\widetilde{c}_{l}^{li}$ and $\widetilde{c}_{i}^{li}$ are the scaled equilibrium concentrations of the liquid and the $i$ solid across the interface, and $\widetilde{\mu}^{li}$ is the scaled equilibrium chemical potential at that interface (we omit $V_{m}$ in what follows since it cancels with the molar normalization on the left-hand side). Differentiating Eq.~(\ref{eq:GLM_PF}) along the equilibrium phase boundary at $T_{p}$ and using Eq.~(\ref{eq:Li_thermo}) gives
\begin{equation}
\begin{aligned}
L_{i}
&=T_{p}\,\Delta c_{\alpha}\,\left.\frac{\partial\widetilde{\mu}}{\partial T}\right|_{p,\text{eq}}\!\bigl(\widetilde{c}_{l}^{li}-\widetilde{c}_{i}^{li}\bigr)\,H\nonumber\\
&=T_{p}\,\Delta c_{\alpha}\,\left.\frac{\partial\widetilde{\mu}}{\partial c}\right|_{p,\text{eq}}\!\left.\frac{\partial c}{\partial T}\right|_{p,\text{eq}}\!\bigl(\widetilde{c}_{l}^{li}-\widetilde{c}_{i}^{li}\bigr)\,H,
\label{eq:latent_chain}
\end{aligned}
\end{equation}
where the second line applies the chain rule. The two factors on the second line are evaluated as follows. Along the linearized liquidus on the $i$ side of $T_{p}$, $T=T_{p}+m_{i}(c-c_{p})$ with slope $m_{i}<0$, so
\begin{equation}
\left.\frac{\partial c}{\partial T}\right|_{p,\text{eq}}=\frac{1}{m_{i}},
\qquad
\text{i.e.,}\qquad
\left|\frac{\partial c}{\partial T}\right|_{p,\text{eq}}=\frac{1}{|m_{i}|}.
\label{eq:dcdT}
\end{equation}
For the chemical-potential slope, the definition $\widetilde{c}=(c-c_{p\beta})/\Delta c_{\alpha}$ gives $\partial\widetilde{c}/\partial c=1/\Delta c_{\alpha}$, so Eq.~(\ref{eq:lambdaTilde_def}) implies
\begin{equation}
\left.\frac{\partial\widetilde{\mu}}{\partial c}\right|_{p,\text{eq}}=\frac{1}{\Delta c_{\alpha}}\,\left.\frac{\partial\widetilde{\mu}}{\partial\widetilde{c}}\right|_\text{eq}=\frac{\widetilde{\lambda}}{\Delta c_{\alpha}}.
\label{eq:dmu_dc}
\end{equation}
Substituting Eqs.~(\ref{eq:dcdT}) and~(\ref{eq:dmu_dc}) into Eq.~(\ref{eq:latent_chain}) and using $\widetilde{c}_{l}^{li}>\widetilde{c}_{i}^{li}$ to keep $L_{i}$ positive, the $\Delta c_{\alpha}$ factors cancel and we arrive at
\begin{equation}
{\;L_{i}=\frac{T_{p}\,\widetilde{\lambda}}{|m_{i}|}\bigl(\widetilde{c}_{l}^{li}-\widetilde{c}_{i}^{li}\bigr)\,H.\;}
\label{eq:latentlambda}
\end{equation}

Equation~(\ref{eq:latentlambda}) provides one equation per interface relating $\widetilde{\lambda}$ to the latent heat. Using $L_{\alpha}=15.9$~kJ/mol and $L_{\beta}=28.3$~kJ/mol from Route~1, the parameters in Supplementary Table~\ref{tabs:constant}, and the equilibrium concentrations $\widetilde{c}_{l}^{l\alpha}$, $\widetilde{c}_{\alpha}^{l\alpha}$, $\widetilde{c}_{l}^{l\beta}$, $\widetilde{c}_{\beta}^{l\beta}$ read directly from the Folch--Plapp phase diagram at $T_{p}$, Eq.~(\ref{eq:latentlambda}) yields $\widetilde{\lambda}\approx 25$ from the $\alpha$--liquid interface and $\widetilde{\lambda}\approx 51$ from the $\beta$--liquid interface. As a compromise that resolves both interfaces with a single prefactor and keeps the dimensionless equations well conditioned, we adopt $\widetilde{\lambda}_\text{ref}=30$ together with $H_\text{ref}=1.462\times10^{8}$~J/m$^{3}$ and $W_\text{ref}=2$~nm as a reference parameter set; the product $WH=0.2924$~J/m$^{2}$ then reproduces the dimensional surface energies quoted above.

\paragraph{Width adaptation across superheatings.}
Because all dimensional surface energies satisfy $\gamma_{jk}=\widetilde{\gamma}_{jk}\,WH$, the product $WH$ uniquely sets the surface tension: doubling $W$ and halving $H$ leaves $\gamma_{jk}$ unchanged. The latent heat $L_{i}\propto\widetilde{\lambda}H$ is preserved if $\widetilde{\lambda}$ is rescaled inversely with $H$, i.e.\ $\widetilde{\lambda}\to\widetilde{\lambda}(H_\text{ref}/H)$. The simulated physics, including the front velocity $v$, the liquid-film thickness $h$, and the tip radius $R$, is therefore invariant under this coupled rescaling. We exploit this freedom to enlarge $W$ at low driving forces, where the underlying physical length scales are intrinsically larger. For each superheating $\Delta T$ we use
\begin{equation}
W(\Delta T)=\frac{200~\mathrm{K}}{\Delta T}\times 2~\mathrm{nm}=\frac{400}{\Delta T}~\mathrm{nm},
\label{eq:Wadapt}
\end{equation}
which recovers $W=2$~nm at the reference $\Delta T=200$~K and grows $W$ as $\Delta T$ decreases. This choice maximizes the dimensionless grid spacing and time step (Eq.~(\ref{eq:dt})) and so accelerates the low-$\Delta T$ simulations while leaving $h$, $v$, and $R$ unchanged to within numerical tolerance.

\clearpage

\supplementarynote{Anisotropic $\boldsymbol{\alpha}$-liquid interfacial free energy}{note:anisotropy}

The dependence of the $\alpha$--liquid interfacial free energy on the orientation of the local interface normal is incorporated only at the $\alpha$--liquid interface; the $\beta$--liquid and $\alpha$--$\beta$ interfaces remain isotropic. The dimensionless anisotropy function takes the standard fourfold cubic-harmonic form~\cite{haxhimali2006orientation,zhong2025quantification},
\begin{equation}
a_{s}(\mathbf{n})=1+\epsilon_{1}\!\left(\sum_{i=x,y,z}n_{i}^{\prime\,4}-\frac{3}{5}\right),
\label{eqs:as}
\end{equation}
where $\mathbf{n}$ is the unit normal to the $\alpha$--liquid interface in the lab frame and $n_{i}^{\prime}=\mathbf{n}\cdot\hat{\mathbf{e}}_{i}^{\prime}$ is its projection onto the $\alpha$ crystallographic axis $i^{\prime}\in\{x^{\prime},y^{\prime},z^{\prime}\}$ specified by the orthonormal unit vectors $\mathbf{x}^{\prime}=(x_{1},x_{2},x_{3})$, $\mathbf{y}^{\prime}=(y_{1},y_{2},y_{3})$, and $\mathbf{z}^{\prime}=(z_{1},z_{2},z_{3})$. We use $\epsilon_{1}=0$ for the isotropic simulations and $\epsilon_{1}=0.1$ for the weakly anisotropic simulations of the main text.

The anisotropic surface free energy modifies the evolution equation for $\phi_{\alpha}$ in two ways~\cite{tourret2015growth}. First, the characteristic time on the left-hand side of Eq.~(\ref{eq:phi_evolution}) acquires the orientation-dependent factor
\begin{equation}
\tau(\vec\phi)\,\partial_{t}\phi_{\alpha}\;\longrightarrow\;\tau(\vec\phi)\,a_{s}^{2}(\mathbf{n})\,\partial_{t}\phi_{\alpha}.
\label{eq:tau_aniso}
\end{equation}
Second, the Laplacian contribution $\sigma\,\nabla^{2}\phi_{\alpha}$ arising from the gradient-energy term of the functional derivative $\delta\mathcal{F}/\delta\phi_{\alpha}$ is replaced by the divergence form
\begin{equation}
\nabla\!\cdot\!\bigl[a_{s}^{2}(\mathbf{n})\,\nabla\phi_{\alpha}\bigr]\;+\;\sum_{i\in\{x,y,z\}}\partial_{i}\!\left[|\nabla\phi_{\alpha}|^{2}\,a_{s}(\mathbf{n})\,\frac{\partial a_{s}(\mathbf{n})}{\partial(\partial_{i}\phi_{\alpha})}\right],
\label{eq:aniso_div}
\end{equation}
which reduces to $\nabla^{2}\phi_{\alpha}$ when $\epsilon_{1}=0$. All other terms in the evolution equations for $\phi_{\alpha}$, $\phi_{\beta}$, and $\phi_{l}$ are unchanged.

\paragraph{Two-dimensional implementation.}
In 2D, $\mathbf{n}=(\cos\theta,\sin\theta,0)$ where $\theta$ is the angle between $\mathbf{n}$ and the lab $x$-axis, so that the crystal-frame projections are
\begin{equation}
n_{x}^{\prime}=x_{1}\cos\theta+x_{2}\sin\theta,\qquad
n_{y}^{\prime}=y_{1}\cos\theta+y_{2}\sin\theta,\qquad
n_{z}^{\prime}=z_{1}\cos\theta+z_{2}\sin\theta.
\label{eqs:projection}
\end{equation}
Following Tourret~et~al.~\cite{tourret2015growth}, the anisotropic operator Eq.~(\ref{eq:aniso_div}) is expanded as
\begin{equation}
\begin{aligned}
&\nabla\!\cdot\!\bigl[a_{s}^{2}\,\nabla\phi_{\alpha}\bigr]+\sum_{i=x,y}\partial_{i}\!\left[|\nabla\phi_{\alpha}|^{2}\,a_{s}\,\frac{\partial a_{s}}{\partial(\partial_{i}\phi_{\alpha})}\right]\\
&\hspace{1em}=a_{s}^{2}\,\nabla^{2}\phi_{\alpha}+2a_{s}a_{s}^{\prime}\bigl(\partial_{x}\phi_{\alpha}\,\partial_{x}\theta+\partial_{y}\phi_{\alpha}\,\partial_{y}\theta\bigr)+\bigl(\partial_{x}\phi_{\alpha}\,\partial_{y}\theta-\partial_{y}\phi_{\alpha}\,\partial_{x}\theta\bigr)\bigl(a_{s}a_{s}^{\prime\prime}+(a_{s}^{\prime})^{2}\bigr),
\end{aligned}
\end{equation}
with $a_{s}^{\prime}\equiv\partial a_{s}/\partial\theta$ and $a_{s}^{\prime\prime}\equiv\partial^{2}a_{s}/\partial\theta^{2}$ obtained by differentiating Eq.~(\ref{eqs:as}) through the chain rule and Eq.~(\ref{eqs:projection}):
\begin{equation}
\begin{aligned}
\frac{\partial a_{s}}{\partial\theta}&=4\epsilon_{1}\!\left(n_{x}^{\prime\,3}\frac{\partial n_{x}^{\prime}}{\partial\theta}+n_{y}^{\prime\,3}\frac{\partial n_{y}^{\prime}}{\partial\theta}+n_{z}^{\prime\,3}\frac{\partial n_{z}^{\prime}}{\partial\theta}\right),\\
\frac{\partial^{2}a_{s}}{\partial\theta^{2}}&=4\epsilon_{1}\!\left[3n_{x}^{\prime\,2}\!\left(\frac{\partial n_{x}^{\prime}}{\partial\theta}\right)^{\!2}+n_{x}^{\prime\,3}\frac{\partial^{2}n_{x}^{\prime}}{\partial\theta^{2}}+3n_{y}^{\prime\,2}\!\left(\frac{\partial n_{y}^{\prime}}{\partial\theta}\right)^{\!2}+n_{y}^{\prime\,3}\frac{\partial^{2}n_{y}^{\prime}}{\partial\theta^{2}}+3n_{z}^{\prime\,2}\!\left(\frac{\partial n_{z}^{\prime}}{\partial\theta}\right)^{\!2}+n_{z}^{\prime\,3}\frac{\partial^{2}n_{z}^{\prime}}{\partial\theta^{2}}\right],
\end{aligned}
\end{equation}
where the derivatives $\partial n_{i}^{\prime}/\partial\theta$ and $\partial^{2}n_{i}^{\prime}/\partial\theta^{2}$ follow directly from Eq.~(\ref{eqs:projection}). Because $\theta=\tan^{-1}(\partial_{y}\phi_{\alpha}/\partial_{x}\phi_{\alpha})$, its first derivatives are
\begin{equation}
\partial_{x}\theta=\frac{\partial_{xy}\phi_{\alpha}\,\partial_{x}\phi_{\alpha}-\partial_{y}\phi_{\alpha}\,\partial_{xx}\phi_{\alpha}}{|\nabla\phi_{\alpha}|^{2}},
\qquad
\partial_{y}\theta=\frac{\partial_{yy}\phi_{\alpha}\,\partial_{x}\phi_{\alpha}-\partial_{y}\phi_{\alpha}\,\partial_{xy}\phi_{\alpha}}{|\nabla\phi_{\alpha}|^{2}}.
\end{equation}

\paragraph{Three-dimensional implementation.}
In 3D we work directly with the components of the gradient. Defining $p=\partial_{x}\phi_{\alpha}$, $q=\partial_{y}\phi_{\alpha}$, $u=\partial_{z}\phi_{\alpha}$, and $r=|\nabla\phi_{\alpha}|$, the projections become
\begin{equation}
n_{x}^{\prime}=\frac{x_{1}p+x_{2}q+x_{3}u}{r},\qquad
n_{y}^{\prime}=\frac{y_{1}p+y_{2}q+y_{3}u}{r},\qquad
n_{z}^{\prime}=\frac{z_{1}p+z_{2}q+z_{3}u}{r},
\end{equation}
and their derivatives with respect to the gradient components are
\begin{equation}
\begin{aligned}
\partial_{p}n_{x}^{\prime}&=\frac{x_{1}}{r}-\frac{x_{1}p^{2}+x_{2}pq+x_{3}pu}{r^{3}},\\
\partial_{q}n_{x}^{\prime}&=\frac{x_{2}}{r}-\frac{x_{1}pq+x_{2}q^{2}+x_{3}qu}{r^{3}},\\
\partial_{u}n_{x}^{\prime}&=\frac{x_{3}}{r}-\frac{x_{1}pu+x_{2}qu+x_{3}u^{2}}{r^{3}},
\end{aligned}
\end{equation}
with the analogous relations for $n_{y}^{\prime}$ and $n_{z}^{\prime}$. The derivatives of $a_{s}$ with respect to $p$, $q$, $u$ then take the form
\begin{equation}
\frac{\partial a_{s}}{\partial p}=4\epsilon_{1}\!\left(n_{x}^{\prime\,3}\,\partial_{p}n_{x}^{\prime}+n_{y}^{\prime\,3}\,\partial_{p}n_{y}^{\prime}+n_{z}^{\prime\,3}\,\partial_{p}n_{z}^{\prime}\right),
\end{equation}
and analogously for $\partial a_{s}/\partial q$ and $\partial a_{s}/\partial u$. Substituting into Eq.~(\ref{eq:aniso_div}) gives the two anisotropic contributions
\begin{equation}
\begin{aligned}
&\nabla\!\cdot\!\bigl[a_{s}^{2}(\mathbf{n})\,\nabla\phi_{\alpha}\bigr]+\sum_{i\in\{x,y,z\}}\partial_{i}\!\left[|\nabla\phi_{\alpha}|^{2}\,a_{s}(\mathbf{n})\,\frac{\partial a_{s}(\mathbf{n})}{\partial(\partial_{i}\phi_{\alpha})}\right]\\
&\hspace{1em}=a_{s}^{2}\,\nabla^{2}\phi_{\alpha}+2a_{s}\bigl(\partial_{x}\phi_{\alpha}\,\partial_{x}a_{s}+\partial_{y}\phi_{\alpha}\,\partial_{y}a_{s}+\partial_{z}\phi_{\alpha}\,\partial_{z}a_{s}\bigr)+\sum_{q\in\{x,y,z\}}\partial_{q}t_{q},
\end{aligned}
\end{equation}
where
\begin{equation}
t_{x}=|\nabla\phi_{\alpha}|^{2}\,a_{s}\,\frac{\partial a_{s}}{\partial p},\qquad
t_{y}=|\nabla\phi_{\alpha}|^{2}\,a_{s}\,\frac{\partial a_{s}}{\partial q},\qquad
t_{z}=|\nabla\phi_{\alpha}|^{2}\,a_{s}\,\frac{\partial a_{s}}{\partial u}.
\end{equation}

\paragraph{Numerical strategy.}
The 3D update is implemented as two CUDA kernels: the first computes the auxiliary fields $a_{s}$, $t_{x}$, $t_{y}$, $t_{z}$ at every grid point; the second uses these fields to assemble the anisotropic contribution and advance $\phi_{\alpha}$. Boundary conditions are applied to all four auxiliary fields on the faces, edges, and vertices of the simulation domain. This split-kernel scheme avoids redundant computation and is roughly twice as fast as a single-kernel implementation in our tests. Because each $\alpha$ orientation requires its own auxiliary fields, the scheme is memory-intensive for polycrystalline simulations with many orientations; the present simulations use a single $\alpha$ orientation, so memory is not a binding constraint.

\clearpage

\supplementarynote{Calculation of genus and characteristic length}{note:analysis}

The bicontinuous topology of the $\alpha/L$ network is quantified by two complementary measures: the genus $g$ of the $\alpha$--liquid interface, which counts independent handles, and the characteristic length $\lambda$ extracted from the structure factor of the phase field, which measures the typical ligament size. We follow the marching-cubes-based methodology used by Geslin et al. \cite{geslin2019phase}, summarized below.

\paragraph{Genus from a triangulated interface.}
The $\alpha$--liquid interface is extracted from the order parameter $\phi$ as the level set $\phi=0$ using the marching-cubes algorithm~\cite{lorensen1998marching}, which yields a closed triangulated surface. For a closed orientable surface composed of $N_\text{blob}$ disjoint connected components with total genus $g=\sum_{i}g_{i}$, the Euler characteristic of the triangulation obeys
\begin{equation}
\chi \;\equiv\; N_{V} - N_{E} + N_{F} \;=\; 2N_\text{blob} - 2g,
\label{eq:euler}
\end{equation}
where $N_{V}$, $N_{E}$, and $N_{F}$ are the numbers of vertices, edges, and faces of the triangulation. Solving for $g$ gives
\begin{equation}
g \;=\; \frac{1}{2}\!\left(\,2N_\text{blob} + N_{E} - N_{V} - N_{F}\,\right).
\label{eq:genus}
\end{equation}

\paragraph{Marching-cubes implementation.}
In the marching-cubes scheme, each grid cube is classified by the signs of $\phi$ at its eight corners, yielding $2^{8}=256$ configurations that map, through a precomputed lookup table, to a fixed set of interior triangles and intersected (``cut'') cube edges. Summing over all cubes $i$ and correcting for sharing between neighboring cubes,
\begin{equation}
N_{F} = \sum_{i} N^{i}_\text{tri},
\qquad
N_{E} = \frac{3}{2}\sum_{i} N^{i}_\text{tri},
\qquad
N_{V} = \frac{1}{4}\sum_{i} N^{i}_\text{CutEdge},
\label{eq:VEF}
\end{equation}
where $N^{i}_\text{tri}$ is the number of triangles inside cube $i$ and $N^{i}_\text{CutEdge}$ is the number of cube edges that the interface crosses. The factor $3/2$ in $N_{E}$ reflects that each triangle contributes three edges, each shared between two triangles, and the factor $1/4$ in $N_{V}$ reflects that each surface vertex is shared by four neighboring cubes. Substituting Eq.~(\ref{eq:VEF}) into Eq.~(\ref{eq:genus}) and simplifying gives
\begin{equation}
g \;=\; N_\text{blob} \;+\; \frac{1}{4}\sum_{i} N^{i}_\text{tri} \;-\; \frac{1}{8}\sum_{i} N^{i}_\text{CutEdge}.
\label{eq:genus_calc}
\end{equation}
The two cube-level quantities $N^{i}_\text{tri}$ and $N^{i}_\text{CutEdge}$ are read directly from the 256-entry lookup tables, and $N_\text{blob}$ is obtained by 26-connectivity component labeling of the $\alpha$ phase.

\paragraph{Characteristic length from the structure factor.}
The characteristic ligament size $\lambda$ is extracted from the structure factor of the order parameter~\cite{geslin2019phase},
\begin{equation}
\begin{aligned}
	S(\boldsymbol{k}) = \frac{\sum_{\boldsymbol{r}}\sum_{\boldsymbol{r'}}e^{-i\boldsymbol{k}\cdot\boldsymbol{r}} \left[ \phi(\boldsymbol{r}+\boldsymbol{r'})\phi(\boldsymbol{r})-\langle \phi \rangle ^2\right] }{N^2(\langle \phi^2 \rangle - \langle \phi \rangle ^2)}
	\label{eq:Sk}
\end{aligned}
\end{equation}
where $N=N_{x}N_{y}N_{z}$ is the number of grid points in the analyzed sub-volume and $\langle\cdot\rangle$ denotes the spatial average over that sub-volume. Spherical averaging over directions gives the isotropic spectrum $S(k)$ with $k=|\boldsymbol{k}|$, and the characteristic wavenumber is defined as the spectrum-weighted mean
\begin{equation}
\bar{k} \;=\; \frac{\int k\,S(k)\,dk}{\int S(k)\,dk},
\qquad
\lambda \;=\; \frac{2\pi}{\bar{k}}.
\label{eq:lambda}
\end{equation}
The front-plane ligament size $\lambda_{0}$ is computed analogously, with the averages in Eqs.~(\ref{eq:Sk})--(\ref{eq:lambda}) restricted to a two-dimensional cross section taken at the dealloying front.

\clearpage

\supplementarynote{Solvability theory of the coupled moving interfaces}{note:solvability}

\begin{figure}[t]
\centering
\includegraphics[width=\textwidth]{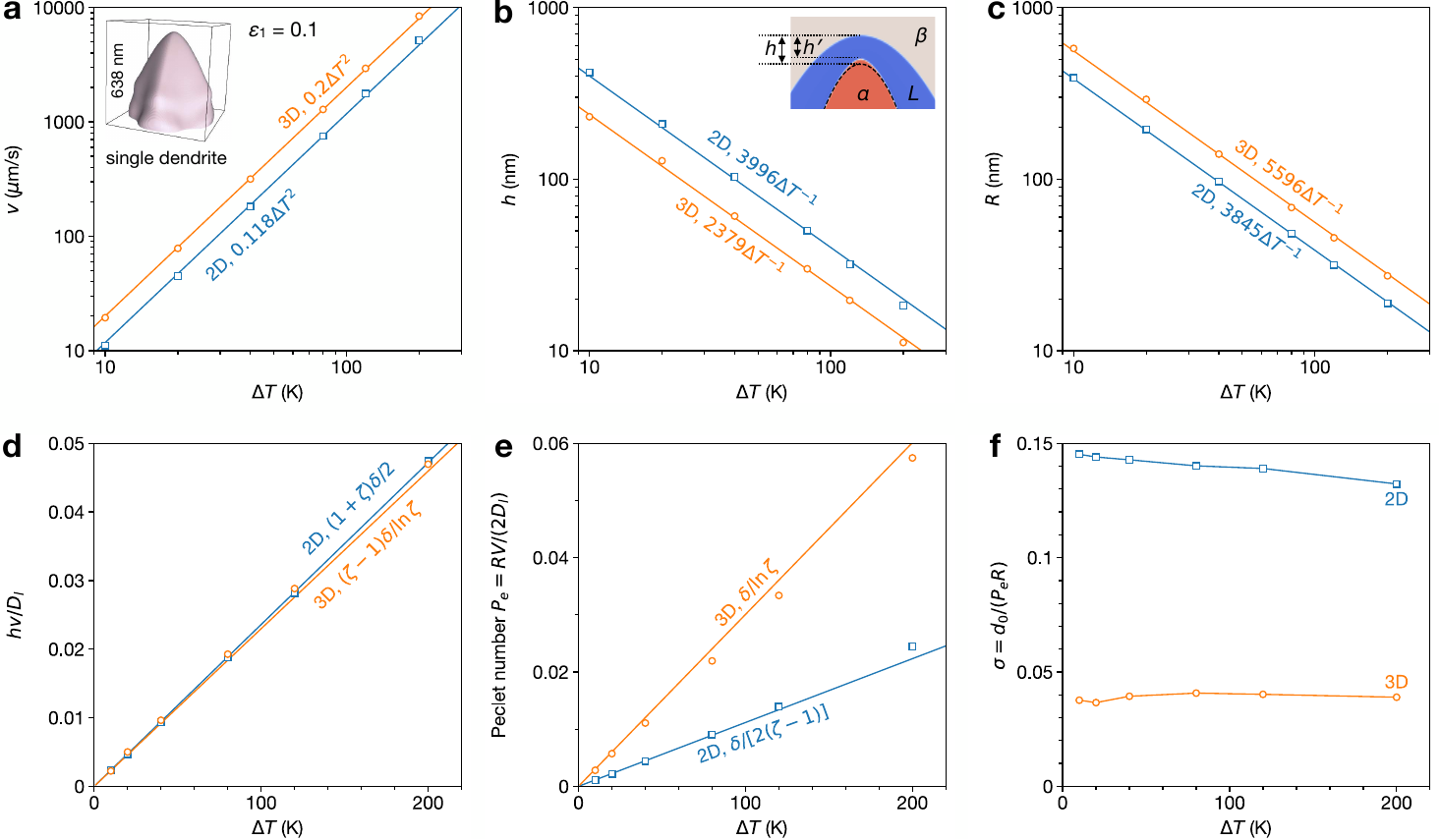}
\caption{\textbf{Steady-state tip kinetics in two- and three-dimensional PF simulations.} Blue squares and orange circles denote 2D and 3D simulations, respectively, of a single $\alpha$ dendrite growing through the Ag-rich liquid film separating it from the parent $\beta$ grain. Here, $D_l$ is the liquid diffusivity and $d_0$ is the capillary length. \textbf{(a)} Growth velocity $v$ versus superheating $\Delta T$, showing $v\propto\Delta T^2$ in both dimensions; the inset shows a steady 3D dendrite at $\Delta T=80$~K. \textbf{(b)} Liquid-film thickness $h$, measured from the fitted parabolic tip to the local $\beta$--liquid interface, versus $\Delta T$; the inset defines $h$ and the directly measured distance $h'$ in 2D. \textbf{(c)} Tip radius $R$ obtained from the parabolic fit versus $\Delta T$. Both $h$ and $R$ scale as $\Delta T^{-1}$. \textbf{(d)} Dimensionless velocity $hv/D_l$ versus $\Delta T$, compared with the analytical predictions from \suppnoteref{note:solvability}: $hv/D_l=(1+\zeta)\delta/2$ in 2D and $hv/D_l=(\zeta-1)\delta/\ln\zeta$ in 3D, where $\delta$ is the driving force and $\zeta$ is a phase-diagram constant. \textbf{(e)} Peclet number $\mathrm{Pe}=Rv/(2D_l)$ versus $\Delta T$, compared with $\mathrm{Pe}=\delta/[2(\zeta-1)]$ in 2D and $\mathrm{Pe}=\delta/\ln\zeta$ in 3D. \textbf{(f)} Microsolvability stability parameter $\sigma=d_0/(\mathrm{Pe}\,R)$ versus $\Delta T$. The nearly constant values, $\sigma\approx0.14$ in 2D and $\sigma\approx0.04$ in 3D, are consistent with solvability theory \cite{barbieri1989predictions}, where $\sigma$ is controlled by the interfacial anisotropy; here $\epsilon_1=0.1$.}
\label{figs:2D3D}
\end{figure}

This note derives the closed-form tip-scale relations [Eqs.~(\ref{eq:hv_Pe})--(\ref{eq:Rv_scaling})] used in the main text and provides the analytical curves plotted in Figure~\ref{fig:5} and Supplementary Figure~\ref{figs:2D3D}. The analysis extends the classical Ivantsov--microsolvability description of free dendritic growth~\cite{ivantsov1947temperature,barbieri1989predictions} to the LFM geometry of PM, in which an advancing $\alpha$--liquid interface and a retreating $\beta$--liquid interface migrate in concert across a thin Ag-rich liquid film. As in the classical theory, the diffusion field around an assumed paraboloidal $\alpha$ tip sets the product $Rv$, i.e.\ the Peclet number $\mathrm{Pe}=Rv/(2D_l)$, together with the film thickness $h$, while microsolvability fixes the stability constant $\sigma=d_0/(\mathrm{Pe}\,R)$ and so closes the selection problem for $R$ and $v$ separately. The new ingredient relative to free growth is that the second moving interface imposes a second concentration boundary condition behind the $\alpha$ tip.

\paragraph{Geometry and governing equation.}
The $\alpha$--liquid tip is approximated as a paraboloid of curvature radius $R$ advancing along $z$ at constant velocity $v$, and the $\beta$--liquid interface as a confocal paraboloid trailing it. In parabolic coordinates normalized to $R$ and centred on the $\alpha$ tip, both interfaces are level surfaces of $\eta$: the $\alpha$--liquid interface lies at $\eta=1$ and the $\beta$--liquid interface at $\eta=\eta_0>1$. In the frame co-moving with the tip, the Ag concentration $c$ satisfies the steady advection--diffusion equation
\begin{equation}
D_l\,\nabla^2 c + v\,\partial_z c = 0.
\label{eq:diff_steady}
\end{equation}
The Gibbs--Thomson condition at each interface,
\begin{equation}
c_{li} = c_p - \frac{T-T_p}{|m_i|} - \frac{\Gamma\,\kappa}{|m_i|},\qquad i=\alpha,\beta,
\label{eq:GT}
\end{equation}
defines capillary lengths $d_i=\Gamma/[|m_i|(c_p-c_{p\alpha})]$; for the Ti--Ag parameters in Supplementary Table~\ref{tabs:constant} ($m_\alpha=-4007.8$~K, $m_\beta=-2618.7$~K), $d_\alpha\simeq 3.2\times10^{-11}$~m and $d_\beta\simeq 4.9\times10^{-11}$~m. For tip radii $R\sim 10$--$10^{3}$~nm, the capillary correction to $c_{li}$ is at most $\sim10^{-3}$, so we ignore the curvature term when solving for the diffusion field and impose $c|_{\eta=1}=c_{l\alpha}$, $c|_{\eta=\eta_0}=c_{l\beta}$, where $c_{l\alpha}$ and $c_{l\beta}$ are read directly from the linearized Ti--Ag phase diagram at the working temperature $T=T_p+\Delta T$. Mass conservation at each interface reads
\begin{equation}
v_n\,(c_p-c_{pi}) = -D_l\,\partial_n c\,\big|_{\eta=\eta_i},
\qquad \eta_\alpha=1,\quad \eta_\beta=\eta_0,
\label{eq:massbal}
\end{equation}
where $v_n$ is the local normal velocity. Equations~(\ref{eq:diff_steady})--(\ref{eq:massbal}) form a closed problem for $c(\eta)$, $\eta_0$, and $\mathrm{Pe}$.

\paragraph{Two-dimensional solution.}
With $x=R\xi\eta$ and $z=R(\eta^2-\xi^2)/2$, the scale factors are $h_\xi=h_\eta=R\sqrt{\xi^2+\eta^2}$. Seeking an Ivantsov-type solution $c=c(\eta)$ reduces Eq.~(\ref{eq:diff_steady}) to
\begin{equation}
\frac{d^2 c}{d\eta^2} + 2\,\mathrm{Pe}\,\eta\,\frac{dc}{d\eta}=0,
\label{eq:ODE_2D}
\end{equation}
with general solution
\begin{equation}
c(\eta)=A\,\mathrm{erf}\!\left(\sqrt{\mathrm{Pe}}\,\eta\right)+B,
\end{equation}
where $A$ and $B$ are fixed by $c(1)=c_{l\alpha}$ and $c(\eta_0)=c_{l\beta}$. Taking the ratio of the two mass balances~(Eq. (\ref{eq:massbal})) at $\eta=1$ and $\eta=\eta_0$ gives
\begin{equation}
\zeta\equiv\frac{c_p-c_{p\alpha}}{c_p-c_{p\beta}}
=\eta_0\,e^{\mathrm{Pe}(\eta_0^2-1)}.
\label{eq:eta0_2D}
\end{equation}
The PF simulations give $\mathrm{Pe} < 0.03$ over the entire range $\Delta T \leq 200$~K (Supplementary Figure \ref{figs:2D3D}e), so $\eta_0\approx\zeta$. Substituting this back into the $\alpha$-interface mass balance and expanding $\mathrm{erf}(x)\approx 2x/\sqrt{\pi}$ yields a linear concentration profile in $\eta$~\cite{brener2005velocity} and the closed-form relations
\begin{equation}
\mathrm{Pe}=\frac{\delta}{2(\zeta-1)},\qquad
h=\frac{1}{2}(\zeta^{2}-1)\,R,\qquad
\frac{hv}{D_l}=\frac{1}{2}(\zeta+1)\,\delta,
\label{eq:result_2D}
\end{equation}
where the dimensionless driving force
\begin{equation}
\delta\equiv\frac{c_{l\alpha}-c_{l\beta}}{c_p-c_{p\alpha}}
\label{eq:delta_def}
\end{equation}
is linear in $\Delta T$ near $T_p$ and the film thickness follows from $h=R(\eta_0^2-1)/2$.

\paragraph{Three-dimensional solution.}
With $x=R\xi\eta\cos\phi$, $y=R\xi\eta\sin\phi$, and $z=R(\eta^2-\xi^2)/2$, axisymmetric solutions of Eq.~(\ref{eq:diff_steady}) again reduce to a one-dimensional ordinary differential equation,
\begin{equation}
\frac{1}{\eta}\frac{d}{d\eta}\!\left(\eta\,\frac{dc}{d\eta}\right) + 2\,\mathrm{Pe}\,\eta\,\frac{dc}{d\eta}=0,
\label{eq:ODE_3D}
\end{equation}
with general solution
\begin{equation}
c(\eta)=A\,E_1(\mathrm{Pe}\,\eta^{2})+B,
\qquad
E_1(z)\equiv\int_{z}^{\infty}\frac{e^{-t}}{t}\,dt.
\end{equation}
Imposing the same two interfacial conditions, taking the ratio of mass balances, and using the small-argument expansion $E_1(z)\approx \gamma-\ln z + z+O(z^2)$ give
\begin{equation}
\zeta=\eta_0^{2}\,e^{\mathrm{Pe}(\eta_0^{2}-1)}\approx \eta_0^{2},
\qquad
\delta\approx \mathrm{Pe}\,\ln\zeta,
\label{eq:eta0_3D}
\end{equation}
and hence
\begin{equation}
\mathrm{Pe}=\frac{\delta}{\ln\zeta},\qquad
h=\frac{1}{2}(\zeta-1)\,R,\qquad
\frac{hv}{D_l}=\frac{(\zeta-1)\,\delta}{\ln\zeta}.
\label{eq:result_3D}
\end{equation}

\paragraph{Microsolvability and tip-scale selection.}
Equations~(\ref{eq:result_2D}) and~(\ref{eq:result_3D}) fix the products $hv$ and the Peclet numbers $\mathrm{Pe}$ as functions of $\delta$ and the phase-diagram constant $\zeta$, but, as in the free-dendrite problem, an additional condition is needed to determine $R$ and $v$ separately. Microsolvability theory~\cite{barbieri1989predictions} supplies this through the stability constant
\begin{equation}
\sigma=\frac{d_{0}}{\mathrm{Pe}\,R}=\frac{2D_l\,d_{0}}{v R^{2}},
\qquad
d_{0}=\frac{1}{2}(d_\alpha+d_\beta),
\label{eq:sigma_def}
\end{equation}
which, for given anisotropy strength $\epsilon_{1}$ of the $\alpha$--liquid free energy, takes a value independent of $\delta$ and hence of $\Delta T$. The PF simulations confirm this prediction: $\sigma\approx 0.04$ in 3D and $\sigma\approx 0.14$ in 2D, both essentially constant over the simulated range $\Delta T\in[20,200]$~K (Figure~\ref{fig:5}f and Supplementary Figure~\ref{figs:2D3D}f). Combining Eq.~(\ref{eq:sigma_def}) with the Ivantsov-type relations yields
\begin{equation}
R\,\propto\,\frac{d_{0}}{\sigma\,\delta},
\qquad
v\,\propto\,\frac{D_l\,\sigma\,\delta^{2}}{d_{0}},
\label{eq:Rv_final}
\end{equation}
with $\zeta$-dependent prefactors that differ between 2D and 3D. Since $\delta\propto\Delta T$ near $T_p$, Eq.~(\ref{eq:Rv_final}) reproduces the observed $R\sim h\sim\lambda_{0}\propto\Delta T^{-1}$ and $v\propto\Delta T^{2}$ scalings of Figure~\ref{fig:5}a--d and Supplementary Figure \ref{figs:2D3D}a--d without adjustable parameters.

\clearpage

\supplementarynote{Supplementary Figures}{note:figures}

\begin{figure}[!h]
\centering
\includegraphics[width=0.5\textwidth]{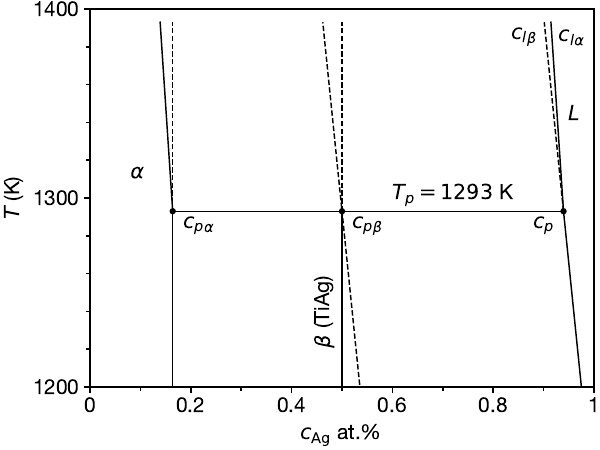}
\caption{\textbf{Phase diagram of the Ti-Ag system around the peritectic temperature $T_p$.} Parameters are given in Supplementary Table \ref{tabs:constant}.}
\label{figs:phase_diagram}
\end{figure}

\begin{figure}[!h]
\centering
\includegraphics[width=0.7\textwidth]{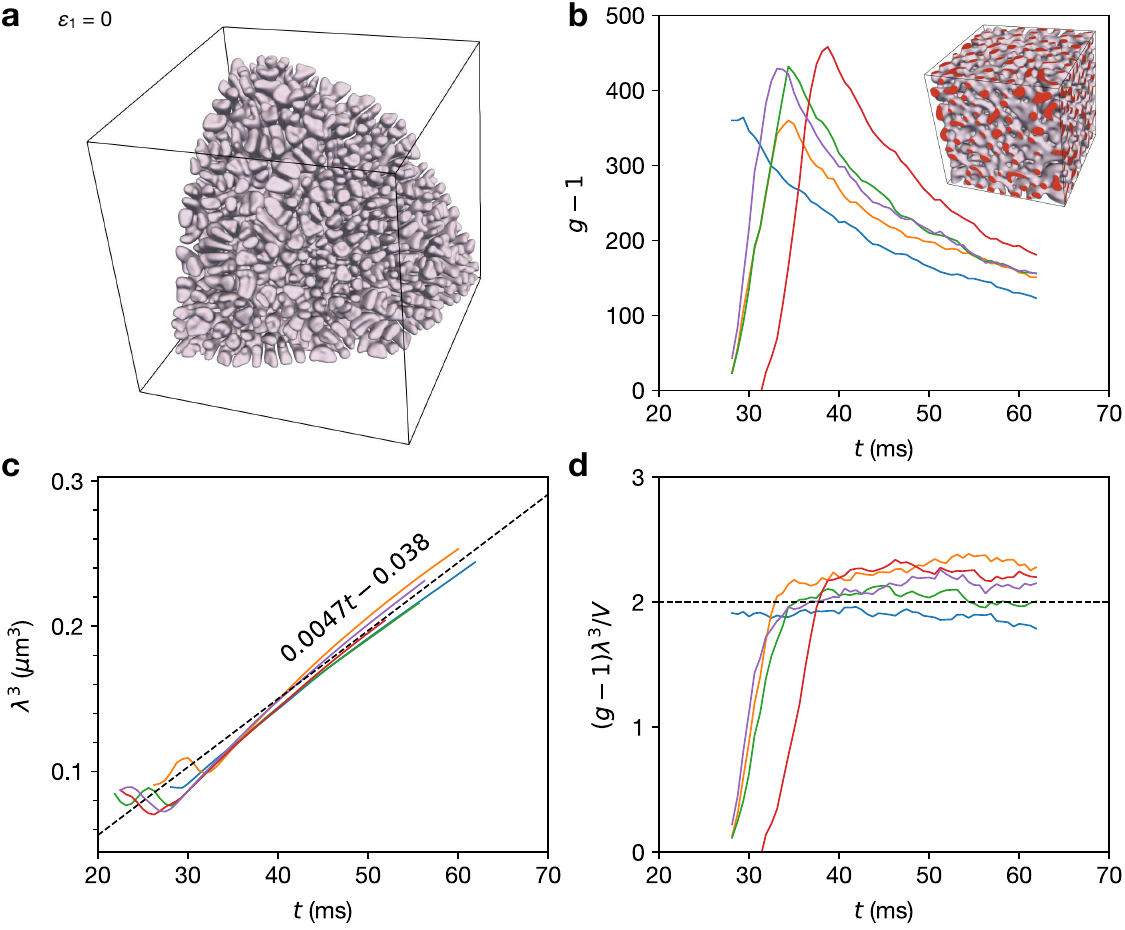}
\caption{\textbf{Topology and coarsening of the bicontinuous $\alpha$/liquid network at $\Delta T=40$~K.} Three-dimensional PF simulation initiated from a Ti-rich $\alpha$ nucleus immersed in the Ag-rich liquid film between neighboring parent $\beta$ grains; curve colors in (b--d) correspond to the analysis sub-volumes in Figure~\ref{fig:3}a. \textbf{(a)} Hyperbranched $\alpha$ network, with the $\alpha$--liquid interface shown in brown and the liquid and $\beta$ phases rendered transparent. \textbf{(b)} Evolution of $g-1$, where $g$ is the genus of the $\alpha$--liquid interface in each sub-volume. Values $g-1\gg1$ confirm that the dealloyed region is topologically bicontinuous, and the subsequent decrease reflects capillarity-driven coarsening; the inset shows the final $\alpha$ morphology. \textbf{(c)} Cube of ligament size, $\lambda^3$, versus time. The curves are shifted in time so as to align $\lambda$ across sub-volumes. The dashed line is a linear fit, giving $k\simeq4.7~\mu\mathrm{m}^3\mathrm{/s}$, consistent with the value at $\Delta T=100$~K (Figure~\ref{fig:3}c). \textbf{(d)} Scaled genus, $(g-1)\lambda^3/V$, with $V$ the sub-volume, saturating near $\sim2$ at long times and indicating robust self-similar coarsening.}
\label{figs:40K}
\end{figure}

\clearpage

\begin{figure}[t]
\centering
\includegraphics[width=\textwidth]{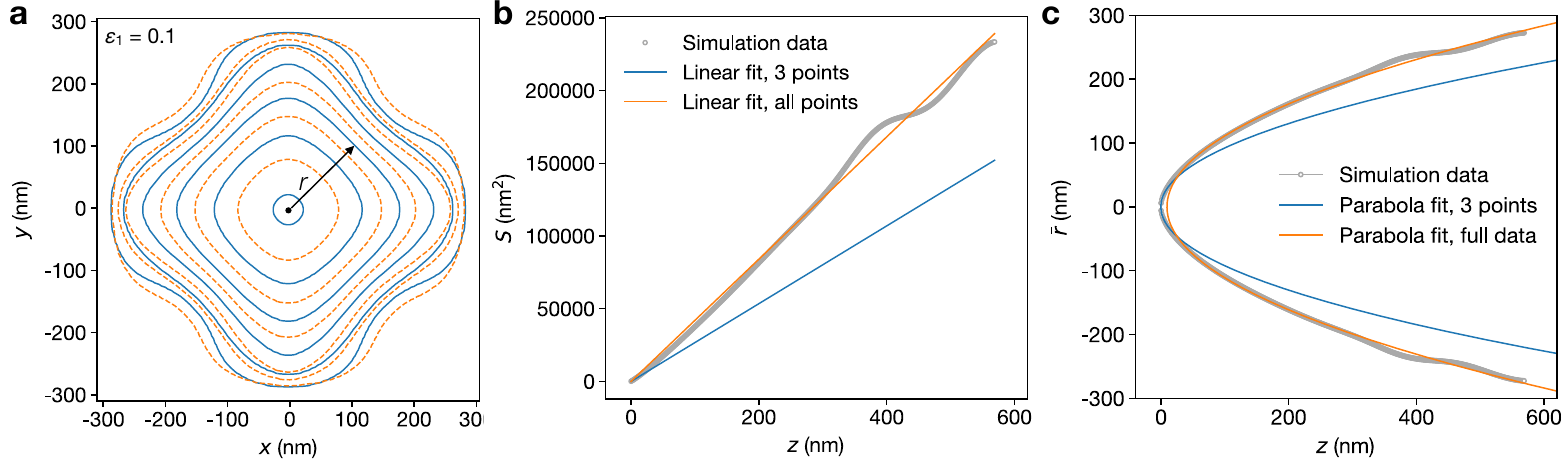}
\caption{\textbf{Parabolic fitting of the $\alpha$--liquid tip.} Analysis of a steady single $\alpha$ dendrite in three dimensions at $\Delta T=80$~K. \textbf{(a)} Contours of the $\alpha$--liquid interface at distances $z$ behind the tip; $r$ is the in-plane radial distance from the contour center. \textbf{(b)} Cross-sectional area $S$ of the $\alpha$ phase versus $z$. For a paraboloidal tip, $S=2\pi Rz$, so the slope gives the tip radius of curvature $R$. \textbf{(c)} Mean cross-sectional radius $\bar{r}=(S/\pi)^{1/2}$ versus $z$, fitted with the parabolic profile using the value of $R$ from (b). The fit determines the parabola-tip position and hence the liquid-film thickness $h$, measured from this fitted tip to the local $\beta$--liquid interface. Fits using only the three points nearest the tip and using all available data give appreciably different values of $R$ and tip position, illustrating the sensitivity of the extracted tip-scale quantities to the fitting range. The fit using all available data is used because it better represents the assumed paraboloidal $\alpha$--liquid interface. }
\label{figs:3Dfit}
\end{figure}

\clearpage

\clearpage

\end{document}